\documentclass{jfm}
\usepackage{graphicx}
\usepackage{epstopdf, epsfig}

\usepackage{hyperref}
\usepackage{color}
\usepackage{amssymb}
\usepackage{amsmath}
\usepackage{setspace}
\usepackage{subfigure}
\usepackage{booktabs}
\usepackage{makecell}
\usepackage[ruled,linesnumbered]{algorithm2e}

\title{A diffusive wetting model for water entry/exit based on the weakly-compressible SPH method}

\author{Shuoguo Zhang\aff{1},
	Yu Fan\aff{1},
	Chi Zhang\aff{1,2},
	Nikolaus Adams\aff{1},
	\and Xiangyu Hu\aff{1}
	\corresp{\email{xiangyu.hu@tum.de}}}

\affiliation{\aff{1}School of Engineering and Design, 
	Technical University of Munich, Garching, 85748, Germany
	\aff{2}Huawei Technologies Munich Research Center , 80992 Munich, Germany
}

\begin{document}
	\begin{spacing}{2.0}
		
	\maketitle
	
	\begin{abstract}
		This paper proposes a diffusive wetting model for 
		the weakly-compressible smoothed particle hydrodynamics (WCSPH) method 
		to simulate individual water entry/exit 
		as well as the complete process from water entry to exit.
		The model is composed of a physically consistent diffusive wetting equation 
		to describe the wetting evolution at the fluid-solid interface,
		a wetting-coupled identification approach 
		to determine the type of fluid particles by 
		taking into account the wetting degree of the contacted solid,
		and a numerical regularization on the fluid particles 
		at fully wetted fluid-solid interface.
		The accuracy, efficiency, and versatility of the present model are validated
		through qualitative and quantitative comparisons with experiments, including the 3-D
		water entry of a sphere, the 2-D water entry/exit of a cylinder, and the complete
		process from water entry to exit of a 2-D cylinder.
		
	\end{abstract}
	

\section{Introduction}
\label{Introduction}
Water entry and exit have been studied for decades and are of great significance for marine
engineering, naval hydrodynamic applications, and more \citep{ZHANG2017Smoo, Watson2021Wat}.
For water entry, in the classical large-scale hydrodynamics perspective based on
\citet{Karman1929Impa} and \citet{Wagner1931Phen}, the inertial effect dominates the impact 
on the free surface. Therefore, factors such as gravity, surface wettability, and air-cushion effect can generally be neglected 
when predicting the hydrodynamic impacting 
force, object trajectory, and induced flow behavior at the initial stage of high-speed 
impact \citep{oliver2002water}. However, as demonstrated by numerous studies \citep{Worthington1897Impa, Albert1951Effect, CHENY1996Extr, Cossali2004role, Ogawa2006Mor},
this simplification is not valid at the later stage, especially when the impacting velocity
is not sufficiently high \citep{kim2019Water, Yoo2022Effect}.
To reveal the unforeseen mechanisms in the physics of impact, \citet{Duez2007Making}
experimentally investigated the relationship between the splashing behavior and surface
wettability and their dependence on impacting velocity. They found that the threshold
velocity for air entrainment is determined by the surface wettability (represented by 
the static contact angle in the experiment). Such a mechanism has been further validated 
and confirmed by experimental studies, in which wettability is modified using different
surface treatments \citep{Gekle2009High, aristoffbush2009Water,
gekle2010Genera, Ueda2012Water, Zhao2014Wett, DIAZ2017eff, Watson2018Jet,
LI2019Experi, Speirs2019Water, Watson2021Wat}.
Compared to the extensive literature on water entry, water exit has been much less investigated
\citep{Zhu2006Wate}. For buoyancy-driven water exit, although no proper theory has been
developed \citep{MOSHARI2014Num}, two typical phenomena have been observed in experiments:
flow separation and free-surface breaking before and after the object breaches the water
surface, respectively \citep{Zhu2006Wate, ZHANG2017Smoo}. It is unclear whether these
phenomena are also influenced by wettability, as in water entry.

Despite the well-established correlation between the splashing behavior and surface wettability 
in experiments, the difficulties in modeling surface wettability make it rarely dealt with 
in practical numerical simulations of water entry \citep{Yoo2022Effect}.
Firstly, because surface wettability is generally governed by many different physical
characteristics (e.g., surface tension, viscous resistance, surface roughness...), the 
high complexity and expensive cost of accounting for all relevant characteristics render
the direct numerical simulation (DNS) impractical, which is similar to 
the dilemma of turbulence simulation.
In particular, while certain physical characteristics, such as viscosity, can be quantified, small-scale but irremissible characteristics, such as surface roughness,
is not feasible to resolve even in large-scale numerical simulations. 
Secondly, although focusing solely on a few dominant physical characteristics offers a
cost-efficient way to characterize surface wettability, this limited consideration will 
still result in significant discrepancies with the experiment. 
For instance, \citet{Yoo2022Effect} employed a DNS model of surface tension to handle 
the surface wettability but predicted a much lower threshold velocity for cavity formation 
than that of \citet{Duez2007Making}.
Furthermore, the dominant characteristics often differ depending on the conditions of 
water entry, so the model built in this compromised way is often limited to specific cases.
Compared to the above mentioned limitations of water entry models, 
the main difficulty in modeling water exit is the lack of mature theoretical support \citep{Zhu2006Wate}.  
Therefore, the existing water exit models are mainly developed with ideal conditions, such as
the inviscid and irrotational flow \citep{korobkin2013inear}.    
Furthermore, as \citet{Oliver2002WaterEA} points out that "...the leading order outer 
problem is linearly stable if and only if the turnover curve is advancing, i.e., the time
reversal of the entry problem is linearly unstable.", simply treating water exit as 
a reversed entry problem, i.e., to apply the water entry model to water exit mechanically, 
is also ill-posed.
Existing numerical simulations of water exit in the literature 
\citep{MOYO2000Free, Zhu2006Wate,  Liu2014mod, ZHANG2017Smoo, LYU2021remo}, 
are not able to accurately reproduce the flow separation and spontaneous 
free-surface breaking in the experiment.
Some researchers have also tried by using larger numerical viscosities 
\citep{SUN2015Nume, ZHANG2017Smoo, LYU2021remo}, 
but the apparent qualitative deviation of the simulation 
from the experiment has not been efficiently improved.
Furthermore, all these open issues in modeling water entry/exit make it currently impossible to simulate 
the complete process from water entry to exit effectively in one model.
Although some attempts have been made in the literature \citep{SUN2015Nume, LYU2021remo, Rosis2022phase},  the state-of-the-art simulations fail to capture not only the typical phenomena of
subsequent water exit, but also the hydrodynamic behaviors of water entry at low-speed impacts.

In this paper, we propose a diffusive wetting model for the WCSPH method to simulate individual water entry/exit 
as well as the complete process from water entry to exit. 
Through a diffusive wetting equation, 
this model utilizes the wetting rate, i.e., the diffusion coefficient, 
to comprehensively characterize the surface wettability 
without introducing complex physical characteristics. 
The resulting progress variable of solid particles quantitatively expresses
the physical wetting degree of the solid.
Together with a wetting-coupled particle identification and a numerical regularization approach,
this model enables the manifestation of the effect of wetting on hydrodynamic behaviors 
in the numerical simulation.
Moreover, by considering the solid surface in the water exit 
as the result of diffusive wetting, 
the proposed model is not only valid for the water exit separately  
but also for both water entry and exit as a complete process.

The remainder of this paper is organized as follows. 
First, Section \ref{WCSPH_method} briefly overviews the Riemann-based WCSPH method and
introduces the coupling between rigid-body and SPH fluid dynamics.
In Section \ref{Wetting_diffusion_model}, the proposed diffusive wetting model is detailed. 
The accuracy, efficiency, and versatility of the present model are qualitatively and
quantitatively validated with several benchmark tests in Sections \ref{Qualitative validations} and \ref{Quantitative validations}, including the 3-D water entry of a sphere,
the 2-D water entry/exit of a cylinder, and the complete process from water entry to exit of
a 2-D cylinder.
Finally, concluding remarks are given in Section \ref{Conclusion}. 
The code accompanying this work is implemented in the open-source SPH library (SPHinXsys)
\citep{Zhang2021CPC} and is available at https://www.sphinxsys.org.

\section{WCSPH method}	
\label{WCSPH_method}
\subsection{Governing equations}
\label{Governing_equations}
Within the Lagrangian framework, the governing equations for an incompressible flow, which is assumed to be isothermal, consist of the continuity and momentum-conservation equations of
\begin{equation} \label{continuity_equation}		
	\frac{d\rho}{dt} = -\rho\nabla\cdot\mathbf{v},
\end{equation}
and
\begin{equation} \label{momentum_conservation_equation}	
	\frac{d\mathbf{v}}{dt} = -\frac{1}{\rho}\nabla p+\nu\nabla^{2}\mathbf{v}+\mathbf{g},
\end{equation}
where $ \rho $ is the density, $t $ the time, $ \mathbf{v} $ the velocity, $ p $ the pressure,
$ \nu $ the kinematic viscosity and $ \mathbf{g} $ the gravitational acceleration. 	

With the weakly-compressible assumption, the system of Eq. \ref{continuity_equation} and Eq. \ref{momentum_conservation_equation} is closed by an artificial isothermal
equation of state (EoS), which estimates the pressure from the density as
\begin{equation} \label{state_equation}
	p=c_0^2(\rho-\rho_{0}), 
\end{equation}
where $c_{0}$ denotes the artificial speed of sound and $\rho_{0}$ the initial
reference density. To restrict the variation in density around 1$\%$
\citep{Morris1997Model}, an artificial sound speed $c_{0} =10U_{max}$ is utilized, with
$U_{max}$ indicating the maximum anticipated flow speed.

\subsection{Riemann-based WCSPH method}
\label{RiemannbasedSPH}
To address the numerical spurious pressure fluctuations in the free-surface flow with violent
impact, 
both the continuity and momentum-conservation equations of Eq.\eqref{continuity_equation} and
Eq.\eqref{momentum_conservation_equation} are discretized by using the Riemann-based WCSPH
method \citep{vila1999particle}, in respect to particle $i$, as following 
\begin{equation} \label{disctretized_continuity_equation}
	\frac{d\rho_{i}}{dt}=2\rho_{i}\sum_{j}\frac{m_{j}}{\rho_{j}}(\mathbf{v}_{i}-\mathbf{v}^{*})
	\cdot\nabla W_{ij}, 
\end{equation}
and
\begin{equation} \label{disctretized_momentum_conservation_equation}
	\frac{d\mathbf{v}_{i}}{dt}=-2\sum_{j}m_{j}(\dfrac{P^{*}}{\rho_{i}\rho_{j}})\nabla W_{ij}+2\sum_{j}m_{j}\frac{\eta\mathbf{v}_{ij} }{\rho_{i}\rho_{j}r_{ij}}\frac{\partial W_{ij}}{\partial r_{ij}}+\mathbf{g},
\end{equation}
where $ m $ is the mass of particle, $ \eta $ the dynamic viscosity, and subscript $ j $ the
neighbor particles. Also, $ \nabla W_{ij} $ denotes the gradient of the kernel function $
 W(|\mathbf{r}_{ij}|,h) $, with $ \mathbf{r}_{ij}=\mathbf{r}_{i}-\mathbf{r}_{j} $ and $ h $
the smooth length. Furthermore, $ \mathbf{v}^{*} = U^{*}\mathbf{e}_{ij}+(\mathbf{\overline{v}}_{ij}-\overline{U}\mathbf{e}_{ij})$, 
where $ \mathbf{e}_{ij}=\mathbf{r}_{ij}/r_{ij} $, 
$\mathbf{v}_{ij}=\mathbf{v}_{i}-\mathbf{v}_{j} $ and $ \mathbf{\overline{v}}_{ij} =(\mathbf{v}_{i}+\mathbf{v}_{j})/2 $ are the relative and average velocities between particles $ i $ and $ j $, respectively. 

Herein, the Riemann solutions $ U^{*} $ and 
$ P^{*} $ of the inter-particle one-dimensional Riemann problem constructed along 
the unit vector $-\mathbf{e}_{ij}$ pointing from particles $ i $ to $ j $ are given by
\begin{equation} 
	\begin{cases} \label{Riemann_solver}
		U^{*}=\overline{U}+\dfrac{P_{L}-P_{R}}{2\overline{\rho}c_{0}}\\[3mm]
		P^{*}=\overline{P}+\frac{1}{2}\overline{\rho}c_{0}(U_{L}-U_{R})\\[3mm]
		(\rho_{L},U_{L},P_{L})=(\rho_{i},-\mathbf{v}_{i}\cdot \mathbf{e}_{ij},p_{i})\\[3mm]
		(\rho_{R},U_{R},P_{R})=(\rho_{j},-\mathbf{v}_{j}\cdot \mathbf{e}_{ij},p_{j})
	\end{cases},
\end{equation}	
where $ \overline{U}=(U_{L}+U_{R})/2 $, $\overline{P}=(P_{L}+P_{R})/2 $, 
and $\overline{\rho}=(\rho_{L}+\rho_{R})/2 $ are inter-particle averages, $ L $ and $ R $ 
the initial left and right states of the Riemann problem.
The utilization of the original intermediate pressure $P^{*}$ in Eq.\eqref{Riemann_solver}
may lead to an excessive dissipation. To mitigate this issue, 
a supplementary low dissipation Riemann solver \citep{article2017Chi}, 
which incorporates a modification on $P^{*}$ while maintaining the intermediate velocity 
$ U^{*} $ in Eq.\eqref{Riemann_solver} unconstrained, reads
\begin{equation} \label{low_dissipation_limiter}
	P^{*}=\overline{P}+\frac{1}{2}\beta\overline{\rho}(U_{L}-U_{R}),
\end{equation}   
where $ \beta = min\:(3\max(U_{L}-U_{R},0),c_{0}) $, representing the limiter, 
is employed in this work.

Furthermore, to tackle the issue of accumulated density error during long-term simulations \citep{Zhang2021CPC} and ensure the numerical stability in free-surface flows, a density reinitialization method proposed by \citet{rezavand2021generalised} is employed, which reinitializes the density field prior to each update in the discretized continuity equation of Eq.\eqref{disctretized_continuity_equation}, as expressed in Eq.\eqref{farfield_density_summation}. Such a scheme has proven effective in mitigating the aforementioned density error and improving the overall accuracy of the numerical scheme.
\begin{equation} \label{farfield_density_summation}
	\rho_{i}=\rho_{0}\dfrac{\sum W_{ij}}{\sum W^{0}_{ij}}+ \max(0, (\rho_{i} -\rho_{0}\dfrac{\sum W_{ij}}{\sum W^{0}_{ij}})) \frac{\rho_{0}}{\rho_{i}},
\end{equation} 
where the superscript $ 0 $ represents the reference value in the initial configuration. Note that the assumption of smooth pressure distribution 
on free-surface particles is applied here due to the weakly compressible assumption.

\subsection{Coupling rigid-body and SPH fluid dynamics}
\label{coupling}
In practical scenarios, the motion of an object in water entry/exit 
cannot be simply described as an ideal rotation-free linear motion along 
the vertical direction, particularly in the later phases of falling 
and rising. Hence, the present model investigates water entry/exit 
under practical conditions by allowing a rigid solid body to freely 
fall and rise without any additional artificial constraints, i.e., 3 
degrees of freedom (DOF) in the 2-D case and 6 DOF in the 3-D case. 
To accurately model the interaction between fluid and solid, 
the coupling of the rigid-body dynamics \citep{sherman2011simbody}
and the SPH fluid dynamics is employed herein.

In detail, SPH firstly conducts a computation of the aggregate force $ F $ 
exerted upon the solid object. This encompasses the fluid pressure force denoted 
as $ F_{total}^{f:p} $, the fluid viscous force designated as $ F_{total}^{f:\nu} $, 
in addition to the gravity $ G $
\begin{equation} \label{total_force}
	F= F_{total}^{f:p} + F_{total}^{f:\nu} +G,
\end{equation}
where the three terms in the right hand of Eq.\eqref{total_force} are respectively defined as 
\begin{equation} 
	\begin{cases} \label{total_force_explanation}
		F_{total}^{f:p}=\sum_{i}f_{i}^{f:p}= -2\sum_{i}\sum_{j}V_{i}V_{j}\dfrac{p_{j}\rho_{i}^{d}+
			p_{i}^{d}\rho_{j}}{\rho_{j}+\rho_{i}^{d}}\nabla W_{ij}\\[3mm]
		F_{total}^{f:\nu}=\sum_{i}f_{i}^{f:\nu}=2\sum_{i}\sum_{j}\nu V_{i}V_{j}\dfrac{\mathbf{v}_{i}^{d}-\mathbf{v}_{j}}{r_{ij}}\frac{\partial W_{ij}}{\partial r_{ij}}\\[3mm]
		G = \sum_{i} m_{i} \mathbf{g}
	\end{cases},
\end{equation}	
where the subscripts $ i $ and $ j $ in present subsection specifically denote solid and 
fluid particles, respectively.
The no-slip boundary condition is imposed at the
fluid-structure interface. Following the fluid-solid coupling scheme in Ref. \citep{Zhang2021multi}, the imaginary pressure $ p_{i}^{d} $, density $ \rho_{i}^{d} $ 
and velocity $ \mathbf{v}_{i}^{d} $ in Eq. \eqref{total_force_explanation} are approximated
as 
\begin{equation} 
	\begin{cases} \label{no_slip_boundary}
		p_{i}^{d}=p_{j}+\rho_{j}r_{ij}max(0,(\mathbf{g}-\frac{d\mathbf{v}_{i}}{dt})\cdot \frac{\mathbf{r}_{ij}}{r_{ij}} )\\[3mm]
		\rho_{i}^{d}=\rho_{0}(\frac{p_{i}^{d}}{\rho_{0}c_{0}^{2}}+1)\\[3mm]
		\mathbf{v}_{i}^{d}=2\mathbf{v}_{i}-\mathbf{v}_{j}		
	\end{cases}.
\end{equation}	
Then, the torque $\mathbf{\tau}$ acting on the center of mass $\mathbf{r}_{cm}$ 
of the falling object is evaluated as
\begin{equation} \label{Torque_equation}
	\tau = \sum_{i}(\mathbf{r}_{i}-\mathbf{r}_{cm})\times (f_{i}^{f:p}+ f_{i}^{f:\nu}+m_{i} \mathbf{g}) .
\end{equation}
With the force $ F $ and torque $\mathbf{\tau}$ in hand, 
the rigid-body dynamics is obtained by solving the Newton–Euler equation
\begin{equation} \label{Newton_Euler_equation}
	\begin{pmatrix} F \\ \tau \end{pmatrix}= \begin{pmatrix} M\mathbf{I} & 0 \\ 0 & \mathbf{I}_{cm} \end{pmatrix}\begin{pmatrix} a_{cm} \\ \alpha \end{pmatrix}+\begin{pmatrix} 0 \\ \omega \times \mathbf{I}_{cm} \omega \end{pmatrix},
\end{equation}
where $ M=\sum_{i}m_{i} $ is the mass of solid object, $ \mathbf{I} $ the identity
matrix, $ \mathbf{I}_{cm} $ the moment of inertia about the center of mass, $ a_{cm} $ 
the acceleration of center of mass, $ \alpha $ the angular acceleration, and $ \omega $ 
the angular velocity. 
All these kinematic values computed by the rigid-body dynamics will be subsequently transmitted to 
the SPH to iteratively update the physical quantities of solid particles, 
including position and velocity, etc \citep{ZHANG2021An}. 
 
\section{Diffusive wetting model}
\label{Wetting_diffusion_model}
\subsection{Diffusive wetting equation}
\label{Wetting_diffusion}
Different from the already wetted surface of a solid object in the typical water exit, 
the wetting of solid-fluid interface in water entry evolves dynamically. 
This evolution includes the wetting spreading on the solid surface and 
the wetting progressing at the solid-fluid interface,
which can be considered as a diffusive process before moisture saturation.  
Consequently, the fully wetted solid surface in water exit can be regarded as the final state 
of the diffusive wetting process. 
Additionally, the wetting rate typically varies with the surface wettability in practice, making it comprehensively characterize the wetting process.

Referring to Fick's second law of diffusion \citep{Fick1855Ueb}, a diffusive wetting 
equation without chemical reactions is proposed as a coarse-grained model 
here to describe this wetting behavior as
\begin{equation} \label{wetting_diffusion_equation}
	\frac{\partial \varphi}{\partial t} = \gamma \nabla^{2}\varphi,
\end{equation}
where the moisture concentration $ \varphi=\varphi(x,t) $ is a function that depends on
location $ x $ and time $ t $, the diffusive wetting coefficient $ \gamma $ represents 
the physical wetting rate with the unit of $ m^{2}/s $. Due to the lack of relevant
experimental data, the experimental-measured $ \gamma $ for each case herein is estimated with the numerical experiment.

In general, in Eq.\eqref{wetting_diffusion_equation}, $ \varphi $ represents the absolute
moisture, defined as the mass of water per unit volume of the solid, with the unit of $ kg/m^{3} $.    
However, in the present SPH model, where a homogeneous solid without any fluid
particles penetrated is considered, it is not an easy task to directly measure the moisture content in 
the unit volume of the solid and predict the concentration based on absolute moisture.    
To conveniently quantify the concentration, the relative moisture $ \varphi^{*}= \varphi/\varphi_{\infty} $ expressed as a percentage is referred to, where $ \varphi_{\infty} $
is the saturated absolute moisture, and then the Eq.\eqref{wetting_diffusion_equation} is rewritten
as 
\begin{equation} \label{relative_wetting_diffusion}
	\frac{\partial \varphi^{*}}{\partial t} = \gamma \nabla^{2}\varphi^{*},
\end{equation}   
where $ \varphi^{*}\in [0,1]$ represents different wetting degrees, for example, 
$ \varphi^{*} $=1 denotes the fully wetted state and $ \varphi^{*} $=0 represents the 
dry state.      
Then the modified diffusive wetting equation Eq.\eqref{relative_wetting_diffusion} could 
be discretized by the SPH method as \citep{CLEARY1998Mode,tang2023integrative} 
\begin{equation}
	\label{discretized_wetting_equation}    	
	\frac{d \varphi^{*}_{i}}{d t} = 2\gamma\sum_{j}\dfrac{m_{j}\varphi^{*}_{ij}}{\rho_{j}r_{ij}}\frac{\partial W_{ij}}{\partial r_{ij}},
\end{equation}
where $\varphi^{*}_{i}$ is the relative moisture of the solid particle, 
and $\varphi^{*}_{ij}=\varphi^{*}_{i}-\varphi^{*}_{j}$ , 
where $\varphi^{*}_{j} \equiv 1$,
the difference between the solid particle and its neighbouring fluid particle.

\begin{figure}
	\centering     
	\includegraphics[width=\textwidth]{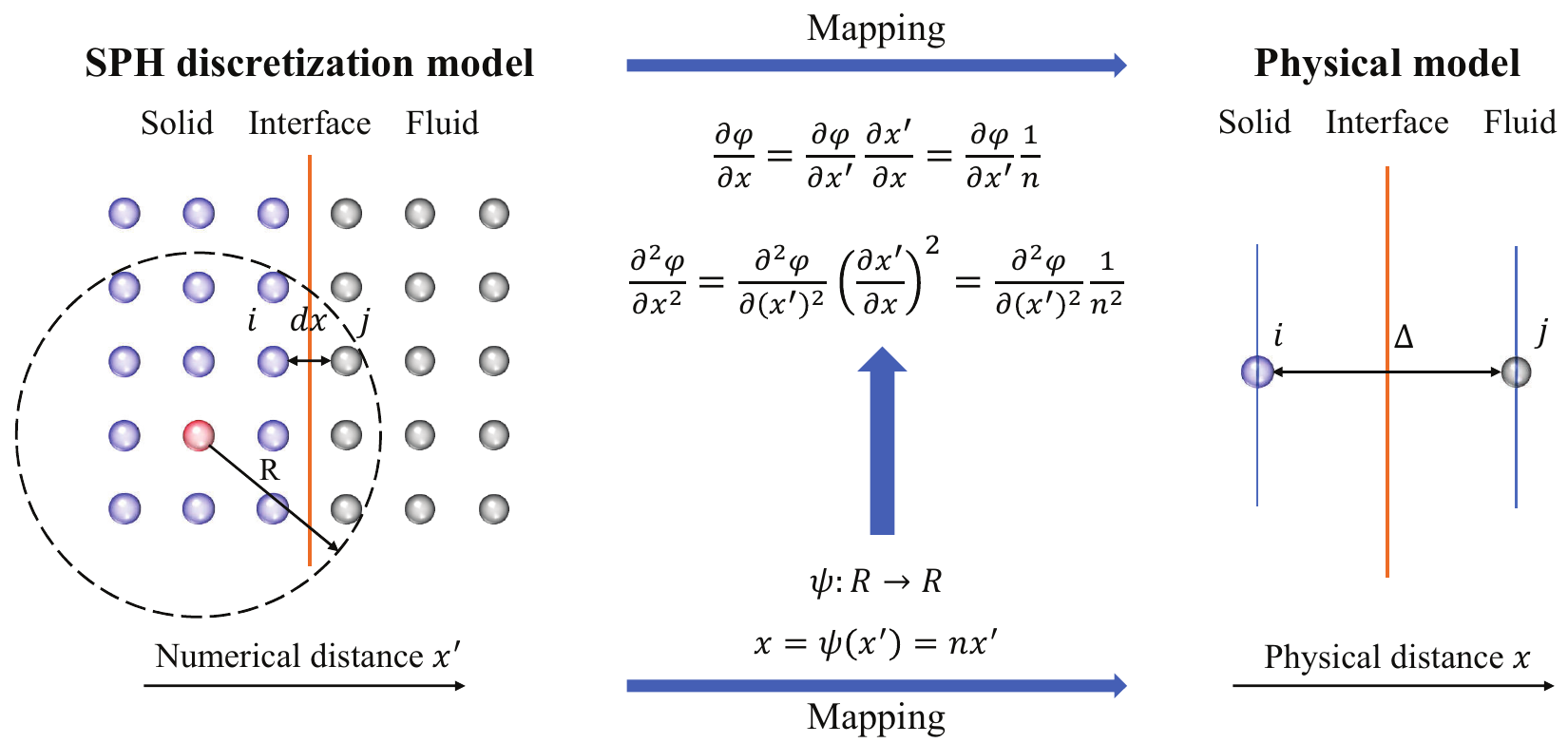}
	\caption{The mapping rule from numerical to physical distances in the diffusive wetting.}
	\label{mapping_rule}
\end{figure}

Note that, the present model only captures the wetting
evolution occurring on the outermost-layer solid particles to assess
the wetting degree of the solid object, which is solely contributed
by the surrounding fluid particles. 
Also noted that, the present SPH model employs a cut-off
radius of $R=2h=2.6dx$, where $dx$ represents the initial particle
spacing. This implies that two layers of surface solid particles are
actually involved in the diffusive wetting, as shown in Figure \ref{mapping_rule}. 
In practice, though, the relative moisture of outermost-layer solid
particles will increase more rapidly due to the contribution from
more neighboring fluid particles, unlike the slower increase in the
relative moisture of the second-layer solid particles.
This ensures the feasibility of using the relative moisture of 
the outermost-layer solid particles as the determinant for
assessment, irrespective of the relative moisture of the
second-layer solid particles.

Furthermore, the microscopic physical thickness of the solid surface 
undergoing diffusive wetting should remain unchanged across different resolutions. 
However, the corresponding numerical thickness, i.e., $ dx $, 
as shown in Figure \ref{mapping_rule}, will decrease as the resolution increases. 
To ensure physical consistency with Eq. \eqref{relative_wetting_diffusion}, 
a mapping from physical to numerical distances is introduced. 
This mapping induces a modified numerical scheme 
using the chain rule, as illustrated in Figure \ref{mapping_rule}.
Thus the discretized diffusive wetting equation Eq. \eqref{discretized_wetting_equation} 
becomes
\begin{equation}
	\begin{cases} \label{rescaled_discretized_wetting_equation}    	
		\frac{d \varphi^{*}_{i}}{d t} = 2\gamma^{*}\sum_{j}\dfrac{m_{j}\varphi^{*}_{ij}}{\rho_{j}r_{ij}}\frac{\partial W_{ij}}{\partial r_{ij}} \\
		\dfrac{\gamma}{\gamma^{*}}=\dfrac{1}{(dx)^{2}} 
	\end{cases},
\end{equation}
where the value 1 is dimensional with the unit of $ m^{2} $.

By employing this mapping rule, 
the resolution independence is achieved, 
which enables the physically consistent diffusive wetting process for arbitrary resolutions. 
Figure \ref{wetting_coupled_identification} illustrates the dynamic wetting results of the solid object 
in typical water entry scenarios,
where different diffusive wetting coefficients $ \gamma=\gamma^{*}/(dx)^{2} $ are applied 
in Eq.\eqref{rescaled_discretized_wetting_equation}. As the flow progresses, the adjacent 
dry solid particles get wetted, causing a gradual increase in the relative moisture. 
Among the three wetting conditions, 
a larger diffusive wetting coefficient leads to a higher overall 
relative moisture of the solid surface at the same instant.
\subsection{Treatments on various wetting states}
\label{Treatment_of_boundary_conditions}
In physics, when the fluid comes into contact with the solid surface, 
the imbalance between adhesion and cohesion acting upon the contacted water molecules 
will initiate the process of wetting and redistribution. 
This process continues until the solid is fully wetted, 
at which point the force imbalance eventually disappears, together with 
redistributed near-surface water molecules.
In the present coarse-grained SPH model,
this molecular redistribution process 
is analogized by the different level of numerical regularization of SPH particles.

Currently, there are two mainstream numerical regularization algorithms in SPH, 
i.e., the particle shifting technique (PST)
\citep{Lind2012Incom, SKILLEN2013163Incom, KHAYYER2017236Comp} and the transport-velocity
formulation (TVF) \citep{Adami2013trans, Zhang2017trans}, 
applied to regularize the SPH particle distribution.
Herein, the TVF scheme is utilized, and the particle advection velocity 
$ \widetilde{\mathbf{v}} $ is expressed as follows
\begin{equation} \label{transport_velocity}
	\widetilde{\mathbf{v}}_{i}(t + \delta t)=\mathbf{v}_{i}(t)+\delta t\left(\frac{\widetilde{d}\mathbf{v}_{i}}{dt}-p_{max}\sum_{j} \frac{2m_{j}}{\rho_{i}\rho_{j}} \frac{\partial W_{ij}}{\partial r_{ij}}\mathbf{e_{ij}} \right).
\end{equation}
Here, the global background $ p_{max} $ is chosen as
$p_{max}=\alpha\rho_{0}\mathbf{v}_{max}^{2}$ with the empirical coefficient 
$ \alpha =7.0$, 
where $ \mathbf{v}_{max} $ is the maximum particle velocity at each advection time step. 
Note that the numerical regularization can effectively eliminate 
the unphysical voids induced by the 
tensile instability in the SPH method, 
which guarantees that the real negative pressure in physics could work well.

Since in the free-surface flow the numerical regularization 
is only carried out for inner fluid particles away from free surface, 
the implementation depends on particle identification, 
which classifies fluid particles into inner and free-surface particles.
If one mimics the free-surface particles with 
the water molecules at the solid-fluid interface before wetting
and the inner fluid particles with the water molecules 
near the fully wetted solid surface,
the above implementation of numerical regularization
can be used together with the diffusive wetting model.
Specifically, the numerical regularization is only 
carried out on the fluid particle near fully wetted solid surface,
which relies on the free-surface identification algorithm detailed in the next section.
\subsection{The coupling of particle identification and diffusive wetting}
\label{transition_of_boundary_conditions}
To identify whether a fluid particle is near fully wetted solid surface, 
the present model primarily adopts 
the spatio-temporal free-surface identification approach \citep{Shuoguo2022free}.
Note that, 
since a relationship between the particle identification rule 
and surface wettability is not provided in the original algorithm,
a free-surface particle is immediately identified as inner one 
once it comes into contact with solid surface.

In order to take into account the surface wettability, 
a wetting-coupling mechanism is introduced here 
to the original identification approach.
It utilizes the relative moisture $\varphi^{*}$ of adjacent solid particles 
as an additional criterion for particle identification. 
In brief, apart from satisfying the position divergence threshold 
required by the original identification, 
the transforming from free-surface to inner particles 
must also meet an additional condition, 
viz, being in contact with at least one fully wetted solid particle.
\begin{algorithm} 
	\label{wetting_coupled_spatio_temparol_algorithm}
	\SetAlgoLined
	\KwData{$ \theta$: particle indicator in previous time step, $ \beta$: particle indicator in current time step, $ \nabla \cdot \mathbf{r}$: position divergence, 
		$ n $: number of fluid particles, 
		$i$: fluid particle index,
		$j$: neighboring fluid particle index of particle $ i $,
		$k$: neighboring solid particle index of particle $ i $,
		$\gamma_{thold}$: threshold of position divergence}	        
	\KwResult{free-surface particles: $  \theta = 1 $, $  \beta = 1 $; inner particles: $ \theta = 0 $, $\beta = 0 $. } 
	
	\textbf{Procedure} Initialization {$ \hfill\triangleright $Execute only once}\\ 
	\For{$i=1$ \KwTo $n$}
	{
		$ \theta_{i} = 1 $\;
	}
	\textbf{Procedure} Wetting-coupled spatio-temporal identification \\
	\For{$i=1$ \KwTo $n$}
	{
		$\nabla \cdot \mathbf{r}_{i} = \sum_{j}\frac{m_{j}}{\rho_{j}} \mathbf{r}_{ij} \cdot \nabla W_{ij}+\sum_{k}\frac{m_{k}}{\rho_{k}} \mathbf{r}_{ik} \cdot \nabla W_{ik}$
		
		\uIf({$ \hfill\triangleright $Wetting coupling}){$ \nabla \cdot \mathbf{r}_{i} > \gamma_{thold}$ \& $ k!=Null $ \& $ \nexists \varphi^{*}_{k}=1 $} 
		{$\nabla \cdot \mathbf{r}_{i} = 0.5\gamma_{thold}$}  
		\ElseIf({$ \hfill\triangleright $Spatio-temporal judgment}){$ \nabla \cdot \mathbf{r}_{i} < \gamma_{thold}$  
			\& $ \theta_{i} != 1 $ \& $\nexists \theta_{j} = 1$}{
			
			$\nabla \cdot \mathbf{r}_{i} = 2\gamma_{thold}$ 
			
		}
	}
	\textbf{Procedure} Update identifications with ensuring\\
	\For{$i=1$ \KwTo $n$}
	{
		$ \beta_{i} =1$\\
		\If{ $ \nabla \cdot \mathbf{r}_{i} > \gamma_{thold}$ \&
			$\nexists \nabla \cdot \mathbf{r}_{j} < \gamma_{thold}$ }
		{
			$\beta_{i} =0$ 
		}
		$\theta_{i} =\beta_{i}$ 
	}
	\caption{The wetting-coupled spatio-temporal identification approach. The main procedures of the approach are: Initialization (lines 1 to 4),  Solver (lines 5 to 13) and Update (lines 14 to 21).}
\end{algorithm} 
The corresponding algorithm  
is summarized in Algorithm \ref{wetting_coupled_spatio_temparol_algorithm}.

\begin{figure}
	\centering     
	\subfigure[$\gamma=\gamma^{*}/(dx)^{2}=0 $.]{
		\begin{minipage}{0.6\linewidth}
			\includegraphics[width=\textwidth]{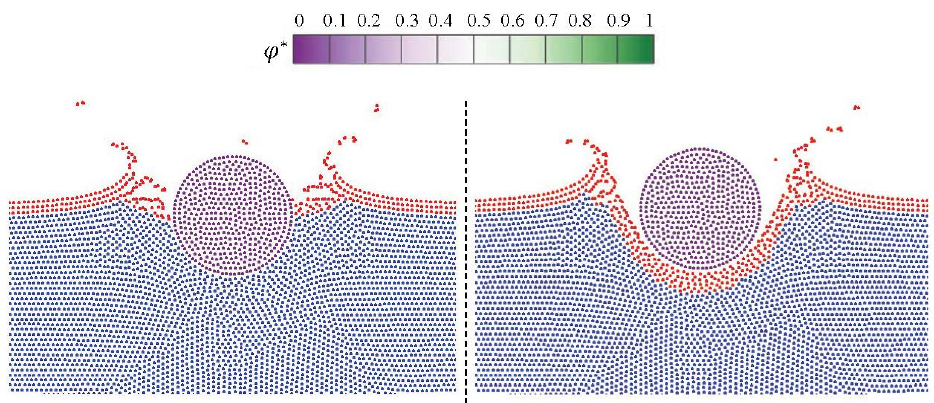}
		\end{minipage}
		\label{gamma_A}
	}	
	\subfigure[
	$\gamma=\gamma^{*}/(dx)^{2}=258.26 m^{2}/s $.]{
		\begin{minipage}{0.6\linewidth}
			\includegraphics[width=\textwidth]{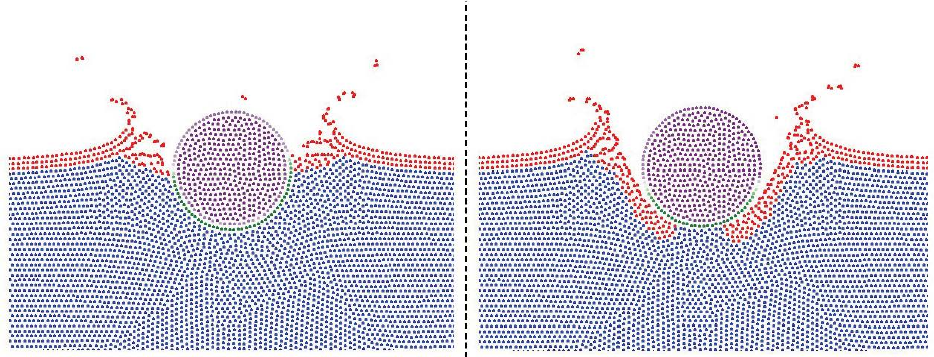}
		\end{minipage}
		\label{gamma_B}
	}
	\subfigure[ $\gamma=\gamma^{*}/(dx)^{2}=2582.64 m^{2}/s $.]{
		\begin{minipage}{0.6\linewidth}
			\includegraphics[width=\textwidth]{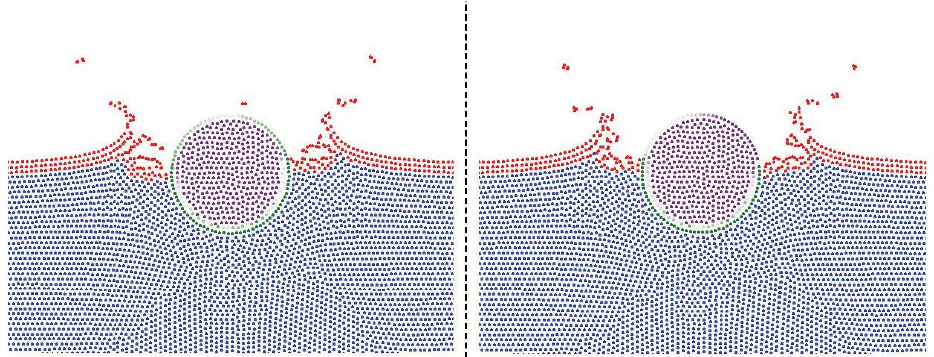}
		\end{minipage}
		\label{gamma_C}
	}	
	\caption{The influence of surface wettability on flow behaviors under three diffusive wetting conditions. Here, the TVF scheme is applied to regularize inner particles, which are respectively identified by the spatio-temporal identification approach (left panel) \citep{Shuoguo2022free} and the wetting-coupled spatio-temporal identification approach (right panel).
		The time instant $ t= 0.27s$. A half-buoyant cylinder with the diameter $ D=0.11m $ is released from $ 0.3m $ above the free surface. The uniform particle spacing is $ dx=D/25 $. The water dynamic viscosity $ \mu $ is $ 8.90 \times 10^{-4} Pa\cdot s $. Fluid particle type: red free-surface particles and blue inner particles.}
	\label{wetting_influence_flow} 
\end{figure}

Since the finite wetting rate in Eq. \eqref{rescaled_discretized_wetting_equation} 
leads to different a delay required for a solid particle to be fully wetted,
consequently, the transforming of inner particle from free-surface one 
will be also delayed. 
\begin{figure}
	\centering 
	\includegraphics[width=\textwidth]{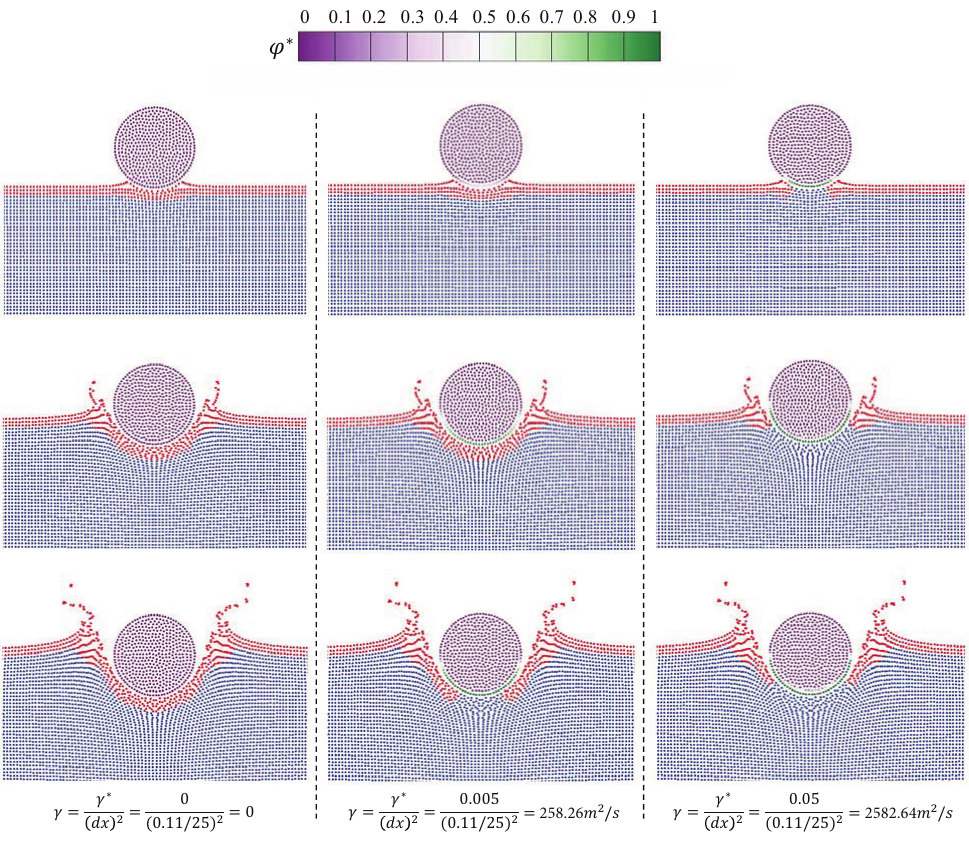}  
	\caption{The delay effect of the wetting-coupled spatio-temporal identification approach 
		under three diffusive wetting conditions. 		
		Here, the TVF scheme \citep{Adami2013trans, Zhang2017trans} is not applied. 
		A half-buoyant cylinder with the diameter $ D=0.11m $ is released from $ 0.3m $ above the free surface. The time instants from top to bottom are $ t=0.23s $, $ 0.25s $ and $ 0.27s $. 
		The uniform particle spacing is $ dx=D/25 $. 
		The water dynamic viscosity $ \mu $ is $ 8.90 \times 10^{-4} \text{Pa}\cdot s $. Fluid particle type: red free-surface particles and blue inner particles. }
	\label{wetting_coupled_identification}
\end{figure}
Figure \ref{wetting_coupled_identification} 
depicts the particle identification at the same instant, 
delayed by the various surface wettabilities.
Subsequently, 
if the numerical regularization is carried out on 
the transformed fluid particles, 
as shown in Figure \ref{wetting_coupled_identification}, by the TVF scheme, 
different hydrodynamic behaviors are obtained, 
as shown in the right panels of Fig. \ref{wetting_influence_flow}.
In comparison, 
if the TVF scheme is implemented based on 
the original particle identification approach, 
the hydrodynamic behaviors are independent of surface wettability, 
as shown in the left panels of Figure \ref{wetting_influence_flow}.

Note that, for a typical water exit problem, 
the submerged cylinder is already fully wetted with $ \varphi^{*}=1$.
Therefore, all the fluid particles near the solid surface are identified 
as inner ones.
Also note that, 
the present identification approach specifically 
allows for the modeling of a complete process from water entry to exit,
as will be shown in Sec. \ref{full_entry_exit},
where the particle identification is fully coupled with 
the dynamical diffusive wetting through the entire process.
\section{Qualitative validations}
\label{Qualitative validations}
\subsection{3-D water entry of a sphere}
\label{3D_water_entry}
\begin{figure}
	\centering     
	\includegraphics[width=0.5\textwidth]{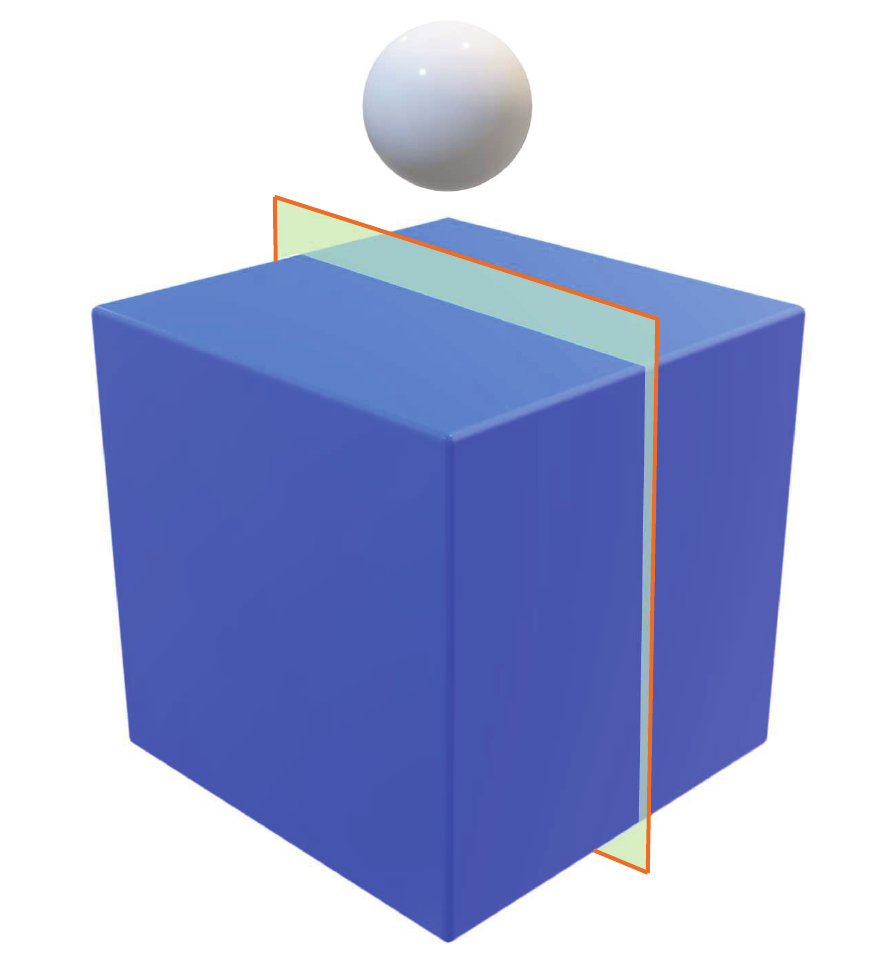}
	\caption{Schematic of the 3D water entry of a sphere with the clipped mid-surface.}
	\label{3D_sphere_schematic}
\end{figure}

The 3-D water entry of a freely falling sphere \citet{Duez2007Making} 
is simulated to qualitatively validate 
the diffusive wetting model to generate various
splash patterns according to the surface wettabilities.
Figure \ref{3D_sphere_schematic} briefly depicts the schematic, where the sphere has a radius
of $D = 0.02m$, an initial relative moisture of $\varphi^{*} = 0$, and a density equivalent 
to that of glass, i.e., $2500kg/m^{3}$. 
The sphere is released at various heights above the
free surface, resulting in different impacting speed of $u_{impct} = 1.4m/s$,
$5m/s$, and $9m/s$. The artificial sound speed $c_{0}$ is defined as 10$u_{impct}$.
A cuboid fluid domain with dimensions of length $L = 3D$, width $W = 3D$, and height 
$H = 3.5D$ is chosen. The dynamic viscosity of water $\mu$ is $8.90 \times 10^{-4} Pa\cdot s$, and its density is $1000kg/m^{3}$. The  gravity acceleration is $g = 9.81m/s^{2}$. 
In all cases, an initial uniform particle spacing of $dx = D/40$ is adopted. Additionally,
to conveniently observe the presence or absence of air entrainment, 
i.e. cavity formation as the water surface closes above the top surface of the sphere,
the mid-surface of 
the fluid domain is clipped, as shown in Figure \ref{3D_sphere_schematic}.

\begin{figure}
	\centering     
	\includegraphics[width=0.6\textwidth]{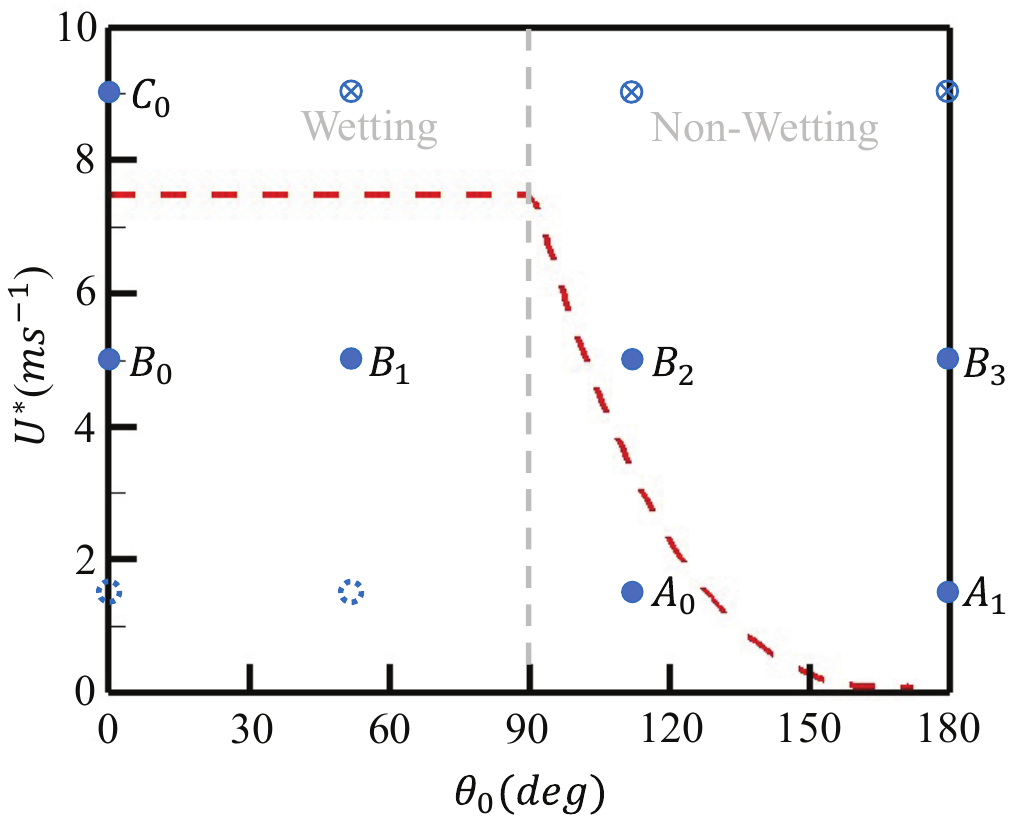}
	\caption{Experimental results of air entrainment as a function of the sphere's threshold velocity 
	$ U^{*} $ and static contact angle $ \theta_{0} $, reproduced from \citep{Duez2007Making}. No air entrainment occurs at the configuration point below the red dotted line, while air cavities of different volumes form above the threshold velocity. Among the 12 configuration points chosen for the present validation, the 7 solid circles represent the ones that are actually tested, while the remaining points, indicated by hollow circles, can be inferred without further investigation.}
	\label{splash_validation}
\end{figure}

\begin{figure}
	\centering     
	\includegraphics[width=\textwidth]{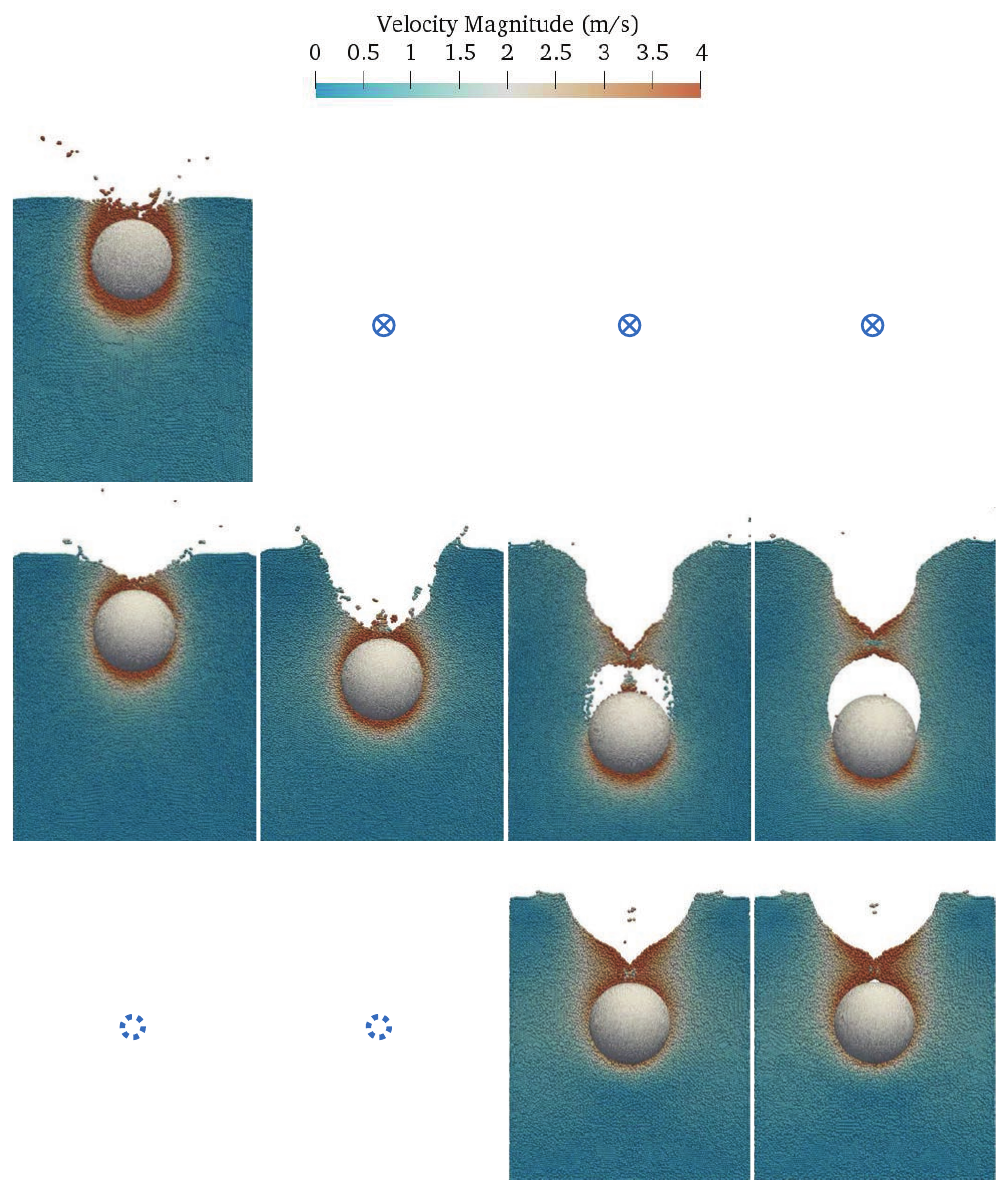}
	\caption{The numerical verification for the air entrainment prediction of \citet{Duez2007Making}.
		Note that, the 7 snap-shots corresponding to 7 simulation setups are arranged according to Fig. \ref{splash_validation}. Note that, for sphere with super-hydrophobic surface, a small volume of air entrainment is generated
		even at the lowest impact speed. Also note that, for super-hydrophilic surface,
		only a small small volume of air entrainment is generated  at the highest impact speed.}
	\label{cavity_prediction}
\end{figure}

Referring to the air entrainment observed during the splashing
processes by \citet{Duez2007Making}, as shown in Fig.
\ref{splash_validation}, we choose 4 wetting rates for 
the 7 tested points. These rates correspond to 4 qualitatively
defined static contact angles representing the
super-hydrophobic, hydrophobic, hydrophilic, and
super-hydrophilic wetting properties of the solid surface,
i.e., $ \gamma=\gamma^{*}/(dx)^{2} = 0$, $25 m^{2}/s$, 
$75 m^{2}/s$, and $\infty$. 
Figure \ref{cavity_prediction} 
gives the air entrainment obtained from the numerical simulations 
corresponding to the experimental setups in as shown in Fig. \ref{splash_validation}.
It is observed that the air entrainment predicted from the simulations 
agree well with experimental observations.
Specifically, 
one can find that a super-hydrophobic sphere 
makes splash with air entrainment or cavity 
(which collapses eventually under the increased ambient pressure)
at all impacting velocities.
However, for a less hydrophobic sphere, 
there is less or no air entrainment for the same impact speed.
With the same wetting properties, 
the splash becomes more evident with a larger volume of air entrainment 
as the impact speed increases.
When the sphere is hydrophilic, 
with the corresponding static contact angle less than 90$^o$, 
much higher impact speed is required to produce an air entrainment. 
Therefore, 
if the impact speed is moderate, 
the ascending splash follows the sphere and quickly accumulates 
at the pole without air entrainment, as shown in Figure \ref{cavity_prediction}.
\subsection{2-D water exit of a cylinder}
\label{Qualitative_validation_water_exit} 
\begin{figure}
	\centering     
	\includegraphics[width=0.7\textwidth]{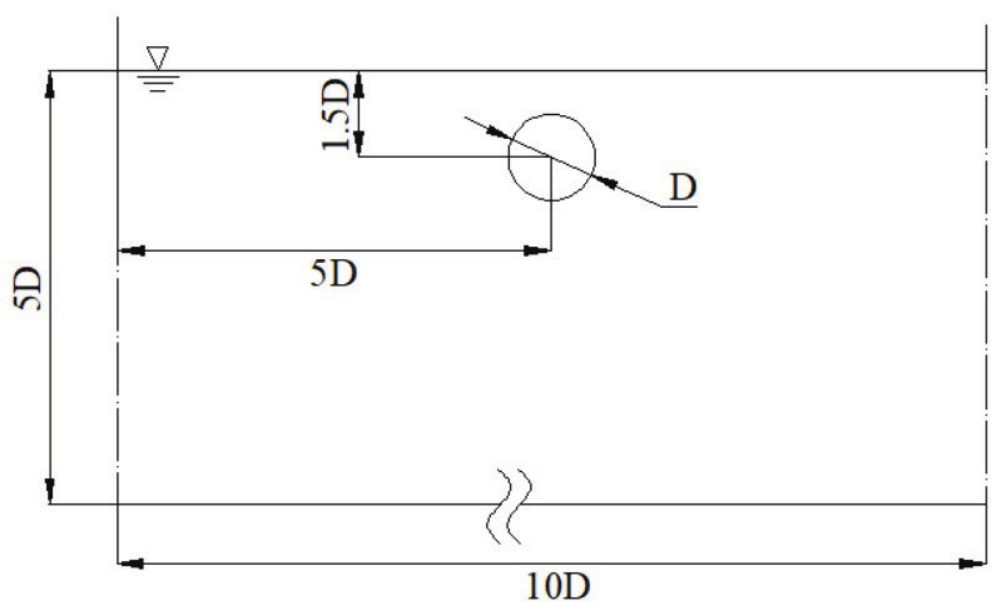}
	\caption{Schematic of the 2-D water exit of a cylinder.}
	\label{schematic_cylinder_exit}
\end{figure}

Following the classical water exit experiment conducted by \citet{Greenhow1983NonlinearFreeSE},
the 2-D water exit of a cylinder is considered 
herein to validate the model's ability 
to capture flow separation and free-surface breaking.   
The schematic of the problem is shown 
in Figure \ref{schematic_cylinder_exit}, 
where a neutrally buoyant cylinder with a diameter of $ D=0.11m $ 
is initially located below the free surface at a distance of $ 1.5D $. 
The submerged cylinder is wetted with an initial relative moisture of $ \varphi^{*}=1 $.
The dimensions of the water tank are $ 5D $ in height and $ 10D $ in width. The water 
dynamic viscosity $ \mu $ and density are $ 8.90 \times 10^{-4} Pa\cdot s $ and 
$ 1000kg/m^{3} $, respectively. 
The artificial sound speed $ c_{0} $ is calculated as 20$\sqrt{5gD}$, where $ g=9.81m/s^{2} $
is the gravity. The cylinder is extracted from the water by a constant force
equal to its weight in the experiment and rises by its buoyancy. To account for this, 
the gravity in the numerical simulation acts only on the fluid, not the freely rising
cylinder. The initial uniform particle spacing is set to $ dx=D/100 $.

\begin{figure}
	\centering    
	\includegraphics[width=0.8\textwidth]{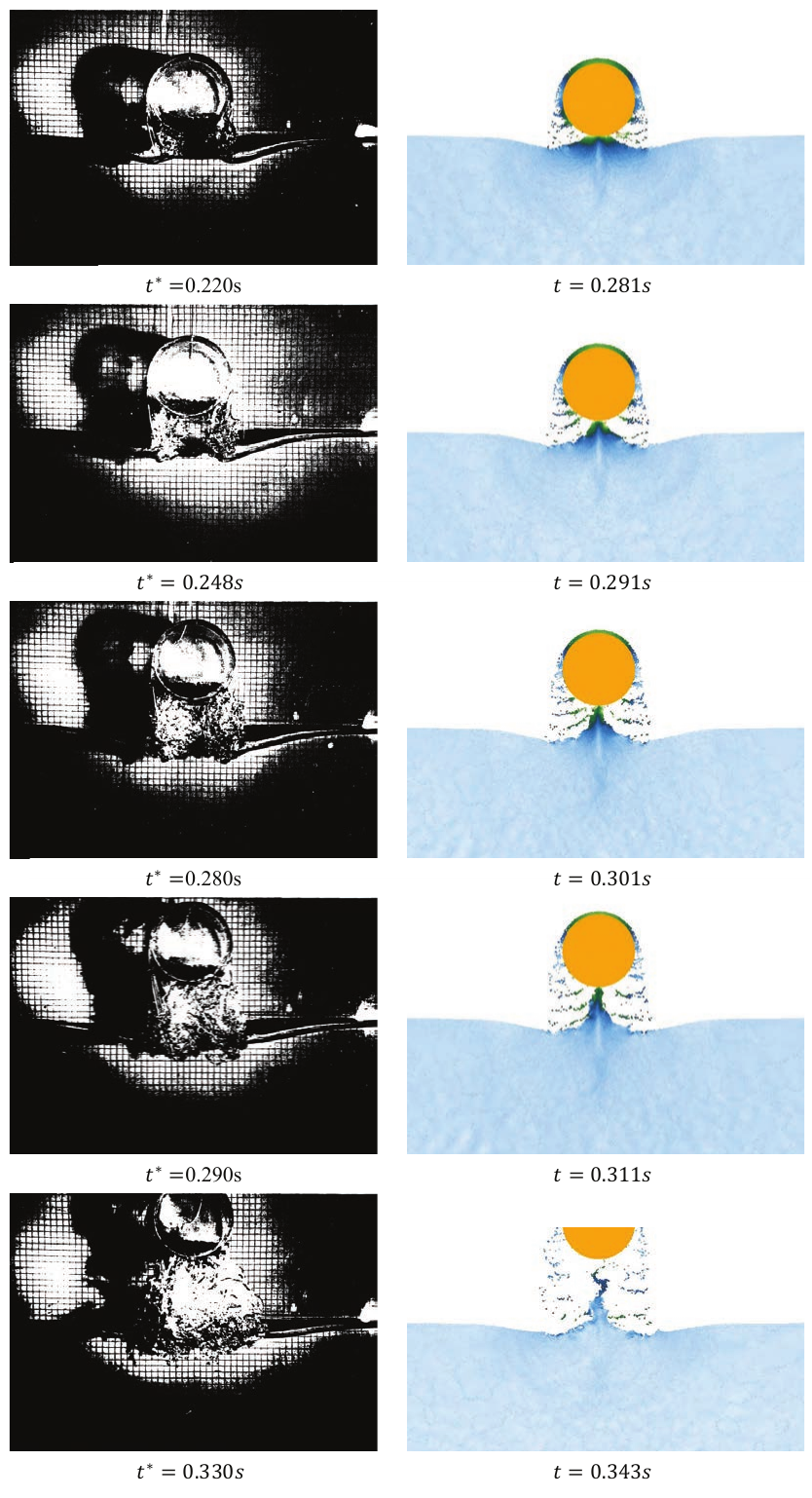}
	\caption{The qualitative comparison of experimental \citep{Greenhow1983NonlinearFreeSE} (left panel) and numerical (right panel) water exit. Note the discrepancies between experimental and simulating instants may be due to the uncertainties.
	The particles are colored with the magnitude of velocity.}
	\label{exp_num_compare}
\end{figure}

\begin{figure}
	\centering     
	\includegraphics[width=\textwidth]{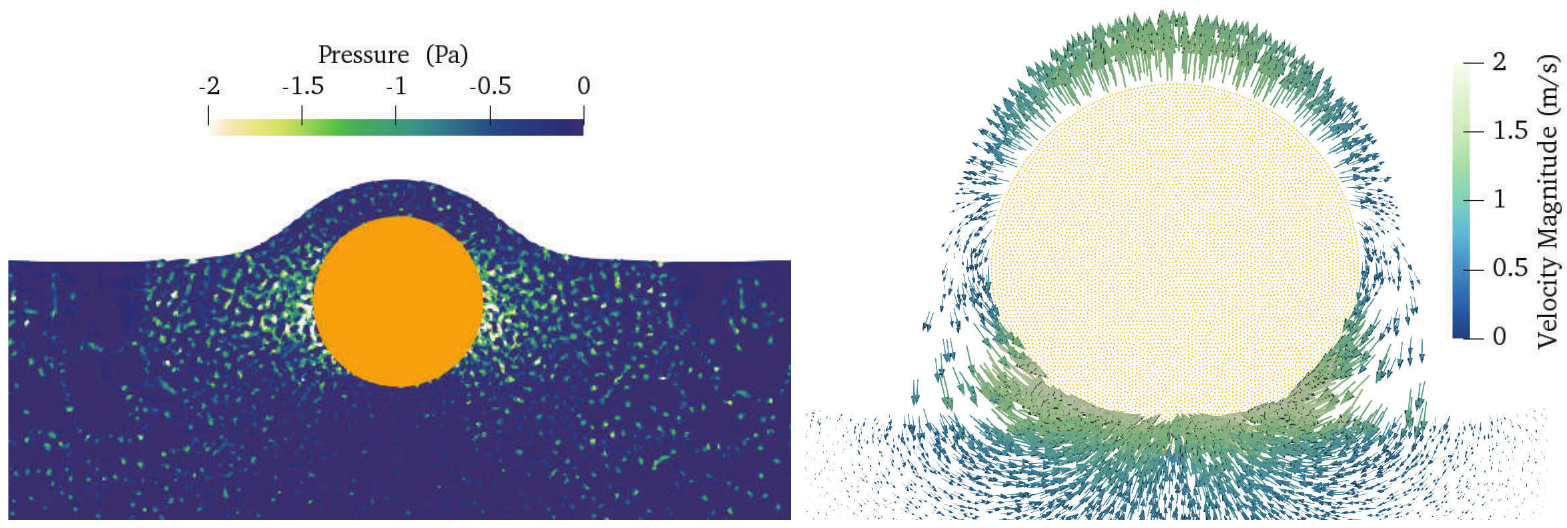}
	\caption{Negative pressure (left panel) and flow separation (right panel).
		The time instant of the pressure contour snapshot is $ t=0.224s $, while that of the flow separation snapshot is $ t=0.263s $ before the happening of free-surface breaking at $ t=0.270s $ in Figure \ref{exp_num_compare}.}
	\label{negative_seperation}
\end{figure}

Figure \ref{exp_num_compare} shows the quite good qualitative comparison 
between the experimental and numerical results at different time intervals.
During the initial phase (from $t=0.185s$ to $0.253s$), 
the water above the cylinder is lifted along with the cylinder, 
resulting in a rapidly downward moving 
and thinning of the water layer. 
Concurrently, a low-pressure region gradually forms on the side of the cylinder, 
with the area and the magnitude of the low-pressure region increasing as the cylinder 
moves upwards \citep{GREEN1988Water}, as shown in the left panel of Figure \ref{negative_seperation}.
When the cylinder is almost leaving the free surface, this phenomenon leads to a 
pressure inversion across the free surface \citep{Greenhow1983NonlinearFreeSE}, 
causing Rayleigh-Taylor
instability \citep{baker1987Rayleigh} and spontaneous free-surface breaking near 
the intersection of the free and cylinder surfaces, also known as "waterfall breaking" \citep{Greenhow1983NonlinearFreeSE}. 
However, this negative pressure will cause unphysical voids to appear in the SPH 
simulation beforehand, so the subsequent spontaneous free-surface breaking has not 
been successfully captured with efficient treatments in most SPH simulations of water 
exit \citep{Buruchenko2017Vali, ZHANG2017Smoo, LYU2021remo}. 
As can be clearly seen that, at the time instance $t=0.270s$, the free-surface breaking 
is realized by the diffusive wetting model without introducing unphysical voids.
This can be attributed to the wetting-coupled spatio-temporal identification approach and 
the particle regularization from the TVF method \citep{Adami2013trans, Zhang2017trans}.
Furthermore, in the right panel of Figure \ref{negative_seperation}, successful capture of 
flow separation before the free-surface breaking is evident.
At approximately $ 110^\circ $ on the rear side of the cylinder, the flow direction of 
the outermost particles deviates significantly from the mean flow and cylinder surface.

In the experiment, when the free surface momentarily breaks,   
thin-layer water in the wake behind the cylinder breaks into droplets \citep{Colicchio2009Expe}.
This remarkable phenomenon is also well reproduced in the present simulation, as shown by the
pronounced scattered falling droplets in the right panels of Figure \ref{exp_num_compare}.
In the following phase (from $t=0.270s$ to $0.343s$), same to the flow behaviors in the
experimental snapshots, the lifted water layer continuously moves downwards along the sides 
of the cylinder but separates from the bulk water due to insufficient downflow velocity. 
Furthermore, it is also important to highlight that as the cylinder breaches the free surface in 
the experiment, the region of low-pressure wake beneath the cylinder pulls a section of the 
free surface downward, creating a depression around the cylinder \citep{Truscott2016Wa}. 
This depression persists throughout the subsequent phase as well, a phenomenon also evident 
in the present simulation.
Hence, the successful reproduction of the complete water exit process, especially the typical
flow separation and spontaneous free-surface breaking, suggests the capability of the present
diffusive wetting model to investigate water exit.

\begin{figure}
	\centering    
	\includegraphics[width=0.8\textwidth]{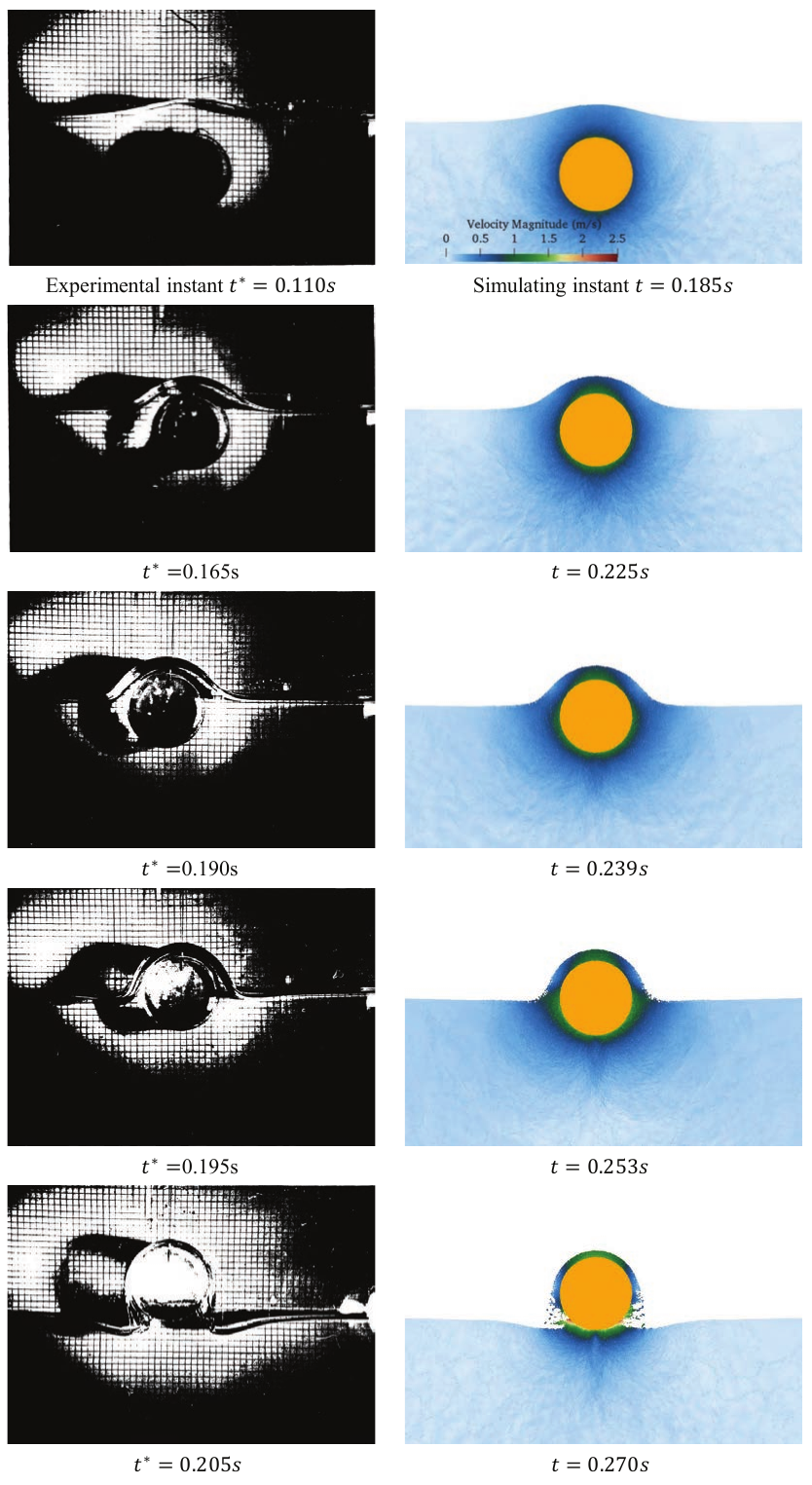}
\end{figure}

\subsection{The complete process from water entry to exit of a 2-D cylinder}
\label{full_entry_exit}
Since the capacity of the present diffusive wetting model in simulating water
entry/exit separately has already been well confirmed through 
the aforementioned cases, 
its potential to simulate the combined processes is further validated herein. 

\begin{figure}
	\centering     
	\includegraphics[width=0.7\textwidth]{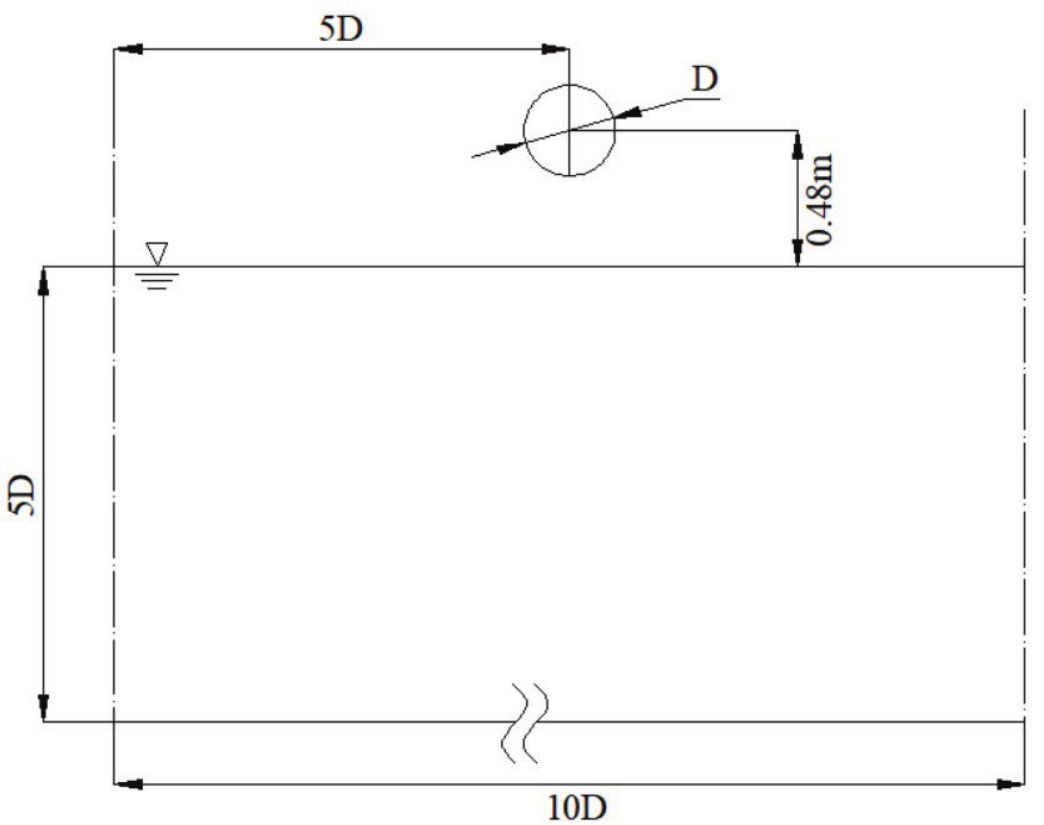}
	\caption{Schematic of 2-D water entry and exit of a cylinder.}
	\label{schematic_cylinder_water_entryexit}
\end{figure}

Here, we consider the model described 
in Section \ref{Qualitative_validation_water_exit}, 
with all parameters kept unchanged except that the cylinder is half-buoyant. 
To obtain an impact speed of  $u_{impct} = 2.89m/s$, 
the cylinder is first lifted by $ 0.48m $ above the free surface and then falls freely, 
as shown in Figure \ref{schematic_cylinder_water_entryexit}.
Three cases with different wetting conditions are considered.
In the first case, 
the cylinder is already wetted ($\varphi^{*}=1$) before impact.
In the second case, 
the cylinder is dry initially with $ \varphi^{*}=0 $,
and the wetting process is controlled by 
the a finite rate ($ \gamma=\gamma^{*}/(dx)^{2}= 0.27m^{2}/s$) 
so that wetting is delayed and fully wetted, 
respectively, during the entry and exit. 
In the last case, the cylinder surface is super-hydrophobic 
so that it keeps dry during the entire process.

\begin{figure}
	\centering    
	\includegraphics[width=\textwidth]{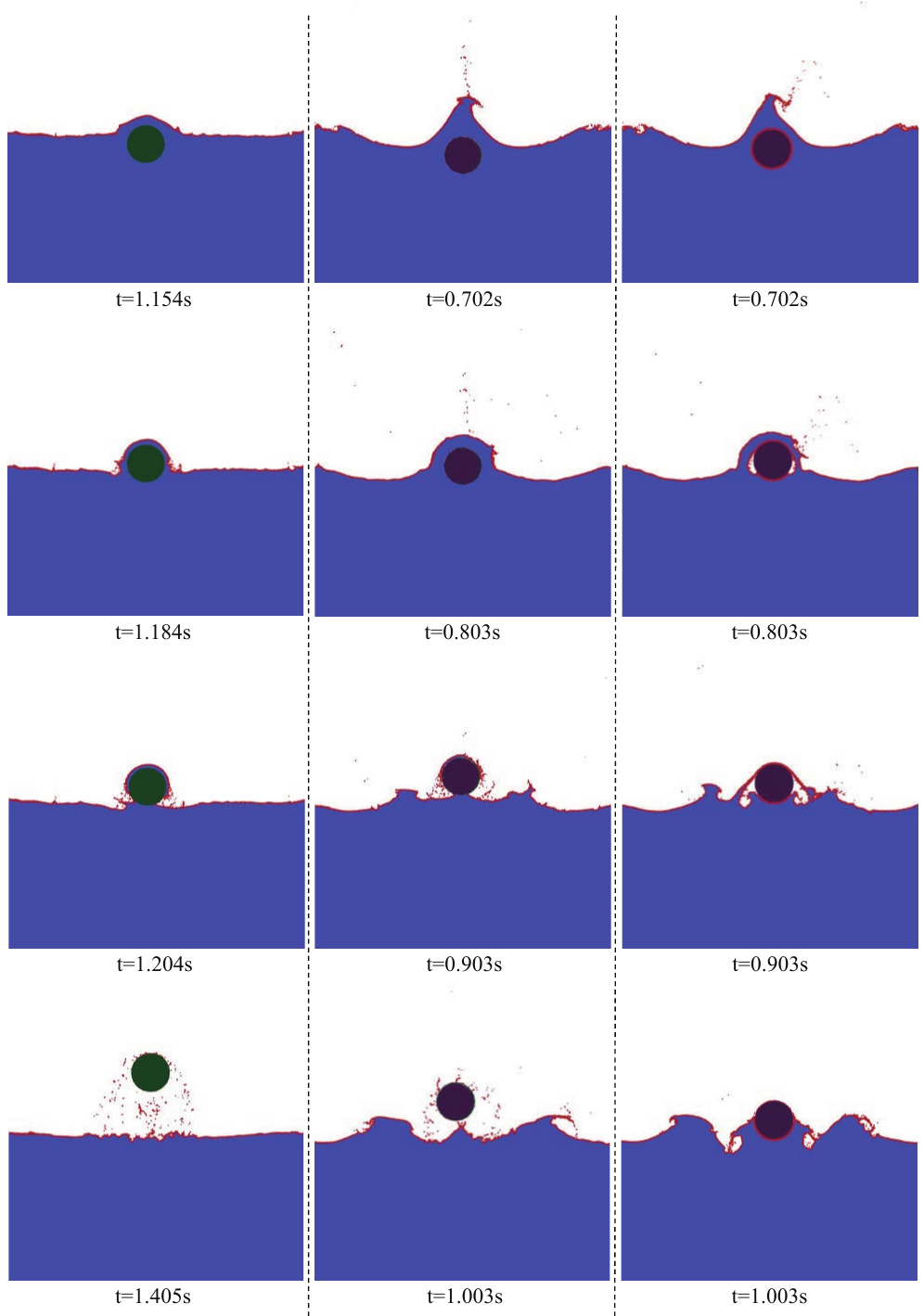}
	\caption{The complete process from water entry to exit in three different wetting conditions. The wetted cylinder (left panel), the dry cylinder with a certain hydrophilicity $ \gamma=\gamma^{*}/dx= 0.27m^{2}/s $ (middle panel) and the superhydrophobic cylinder (right panel).}
	\label{full_process}
\end{figure}

Figure \ref{full_process} presents the snapshots 
with non-wetted fluid particles indicated 
for all the three cases. 
When the cylinder is already wetted before the impact,
as shown in the left panel, 
the fluid particles near the solid surface 
are immediately identified as inner ones and 
imposed with numerical regularization. 
Like the super-hydrophilic sphere as shown in Figure \ref{splash_validation}, 
the cylinder is quickly submerged after the impact 
without generate much splashing.  
After the cylinder descends a significant depth, 
the buoyancy force eventually overtakes the weight and inertial, 
stops the cylinder at about the time instance $t=0.702s$ and raises it up again.
Under the acceleration of buoyancy force, 
the cylinder later leaps out of the water surface 
and reaches a considerable pop-out height 
(the maximum value above the free surface). 
Note that, even with the presence of agitated water surface after impact,
the phenomenon of "waterfall breaking" remains evident,
which is in a good agreement with the water exit described in 
Section \ref{Qualitative_validation_water_exit}. 

In contrast, 
when the cylinder is dry initially, as shown in the middle panel,
the wetting process is delayed during water entry due to 
the finite wetting rate. 
Such delay results in a gradual transformation of 
the near-solid-surface fluid particles into inner ones and 
hence the delayed imposing of the numerical regularization,
which results a cavity with two almost symmetric and vigorous jets. 
During this process, 
the cylinder remains half-submerged
before the retreat flows from both sides cover the cylinder surface.
Note that, 
the maximum descent depth of the cylinder is less than that 
in the previous case, 
attributed to the greater energy dissipation resulting from the jets and splashes.
This also explains the diminished leaping velocity 
and a notable reduction of the pop-up height \citep{Truscott2016Wa} 
when the cylinder breaches water surface again.
Furthermore, during the water exit phase, 
since the cylinder surface is already fully wetted, 
"waterfall breaking" very similar to the previous case is observed.

For the last case with the super-hydrophobic cylinder, 
as shown in the right panel, 
all hydrodynamic behaviors during water entry, 
including the maximum descent depth, 
closely resemble those of the second case.
which aligns with \citet{Duez2007Making}'s prediction.
However, due to the non-wetted surface, the near-solid surface fluid particles 
are consistently identified as non-wetted 
without particle regularization imposed. 
Consequently, the subsequent "waterfall breaking", as seen in 
the previous two cases, does not occur; instead, it is replaced 
by the formation of two cavities on both sides of the cylinder.
Note that the adopted wetting-coupled spatio-temporal particle identification approach 
in the present model ensures that 
these cavities during the water exit are not unphysical voids. 
This phenomenon is similar to the cavitation observed in hydrodynamics.
Interestingly, as the water is further lifted by the rising cylinder, 
a unique thin layer of water resembling a hat remains consistently.
The presence of this hat-like layer increases hydrodynamic resistance, 
leading to a quicker reduction in the upward velocity of the cylinder 
compared to the previous two cases.
As a result, the cylinder does not exhibit 
a distinct leap out of the water but drains the water layer gradually.
\begin{figure}
	\centering    
	\includegraphics[width=\textwidth]{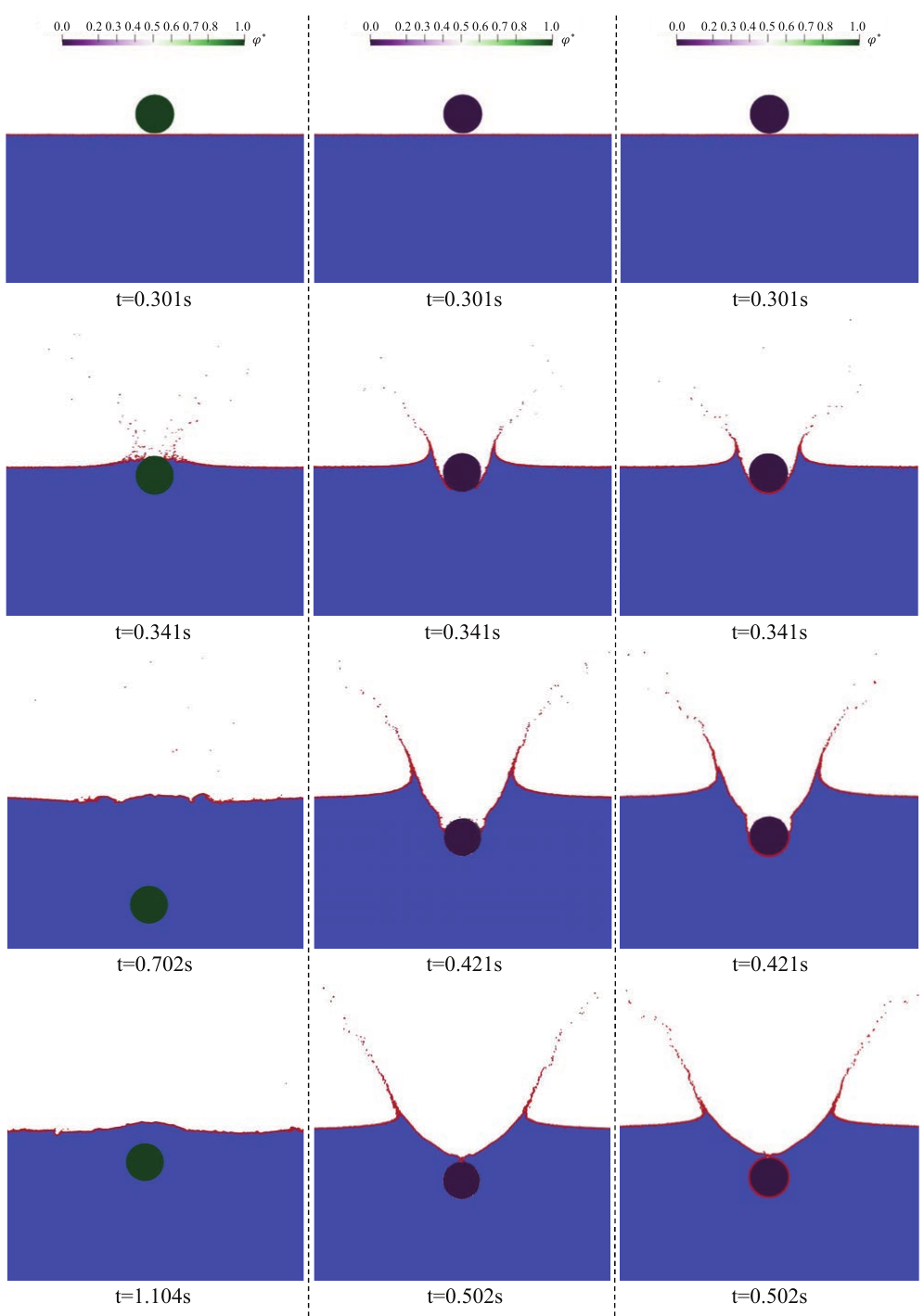}
\end{figure}

\section{Quantitative validations}
\label{Quantitative validations} 
\subsection{2-D water entry of a cylinder}
\label{2D_water_entry} 
In order to increase the reliability of the diffusive wetting model 
in practical application, 
a 2-D water entry of a freely falling cylinder is modeled 
and then quantitatively compared 
with the experiment \citep{Colicchio2009Expe}.
Referring to the experimental setup, 
the diameter and density of a stainless steel circular cylinder 
are given as $ 0.3 m $ and $ 620 kg/m^{3} $ respectively, 
while other geometrical and physical parameters are the same as 
that in Section \ref{full_entry_exit}.  
The poor hydrophilic surface initially is dry and assigned 
with a diffusive wetting rate of $\gamma=\gamma^{*}/(dx)^{2}=0.17m^{2}/s$.

\begin{figure}
	\centering     
	\includegraphics[width=\textwidth]{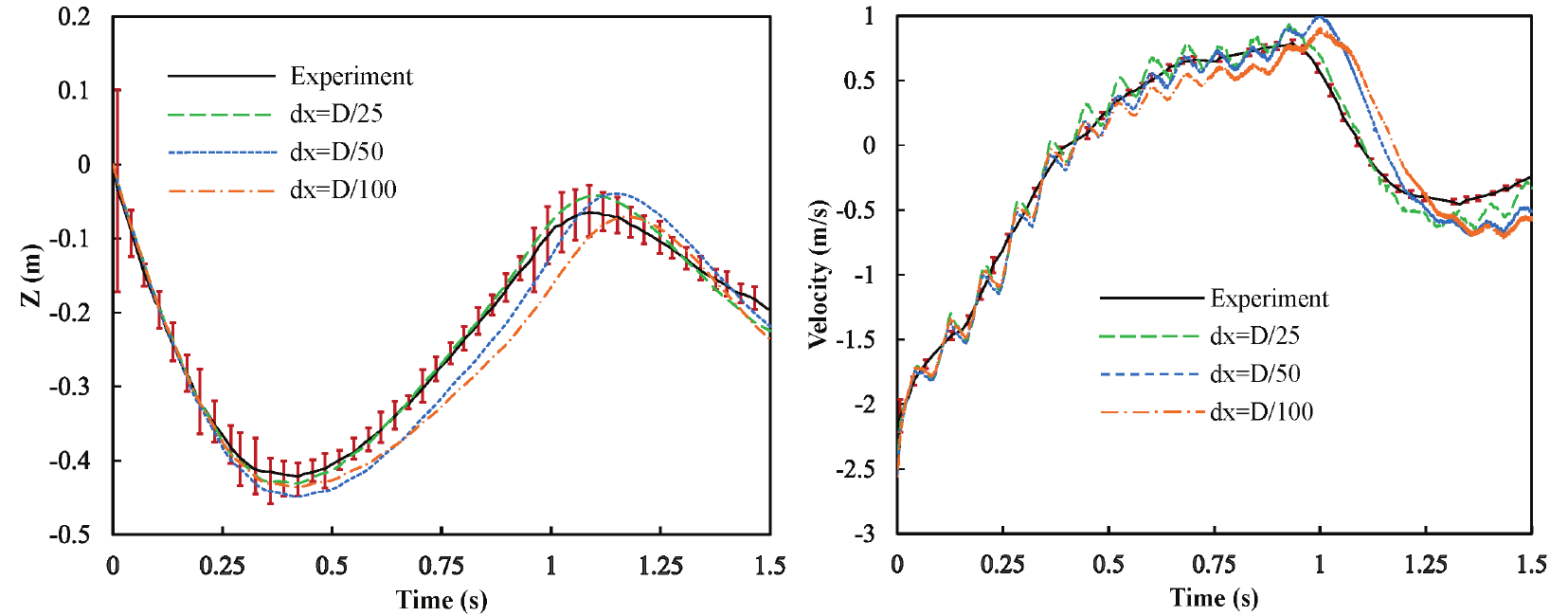}
	\caption{Comparison about the vertical position (left panel) and velocity (right panel) of cylinder center with Experiment \citep{Colicchio2009Expe}.The experimental data are plotted with the standard deviation error bars.}
	\label{position-velocity-comparison}
\end{figure}

The left panel of Figure \ref{position-velocity-comparison} shows the time trace of the
vertical position of the cylinder center throughout the entire process, from water entry 
to exit.
In the early stage of water entry (approximately $ t<0.2s $), the time trace exhibits 
high repeatability with small run-to-run deviations, which agrees well with the experiment.
During the later phase of descent (approximately $ 0.2s<t<0.41s $), the time traces under
different resolutions show a slight divergence, but they remain well within the range 
defined by the standard deviation error bars of the experimental data \citep{Colicchio2009Expe}
and show a convergent tendency.
Moreover, in the present simulation with a finite and small tank size, the water 
wave propagation caused by the splash during water entry will be blocked by the side 
walls, resulting in an elevation of the free surface.
Hence, compared to the experimental time trace in subsequent water exit (approximately 
$ 0.41s<t<1.1s $), the increased water pressure above the cylinder will slow down its ascent 
in the numerical simulation.
In the next subsection \ref{2D_water_exit} about water exit, the initially immersed 
cylinder rises up in a calm water tank without the influence of any violent wave propagation,
and this deviation will be eliminated, which verifies the rationality of the above 
explanation well.

In the right panel of Figure \ref{position-velocity-comparison}, the unsteady hydrodynamic
force \citet{truscott2012Unst} acting on the cylinder will induce the oscillation of 
its vertical velocity throughout the entire process.
Even this, during the stage of water entry, the continuous line, representing the mean
experimental velocity, approximates the fitted curve of the numerical oscillating velocity. 
The vertical velocity during the water exit stage is lower than the experimental value, 
which align with the explained time trace of the vertical position in water exit.

\subsection{2-D water exit of a cylinder}
\label{2D_water_exit}
As the same circular cylinder is used for both experimental water entry/exit \citep{Colicchio2009Expe}, the same cylinder in Section \ref{2D_water_entry} 
is submerged and fully wetted with its center at a depth of $ 0.46m $ 
from the free surface, 
and pushed upwards by the buoyancy force. 
\begin{figure}
	\centering    
	\includegraphics[width=1\textwidth]{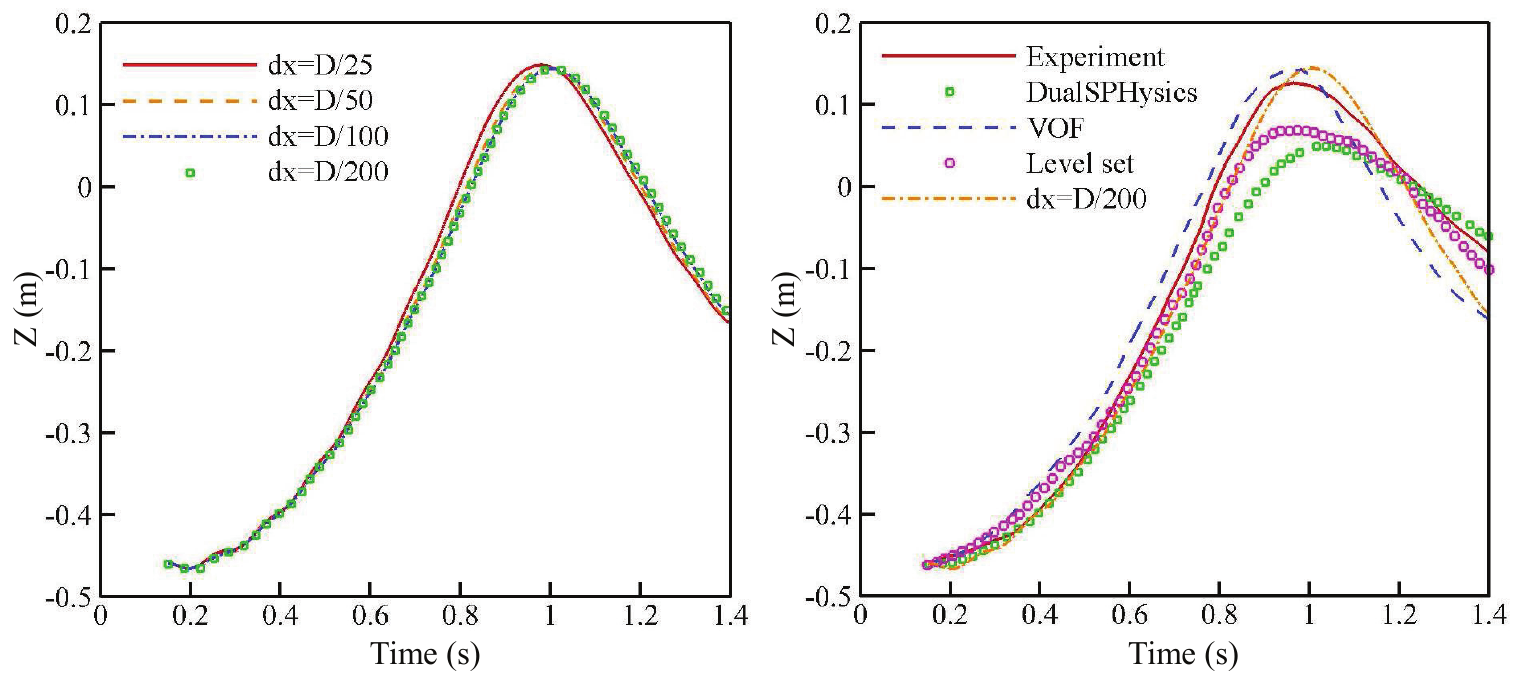}
	\caption{Convergence analysis (left panel) and comparison (right panel) about the vertical position of cylinder center. Experiment \citep{Colicchio2009Expe}, DualSPHysics \citep{Buruchenko2017Vali}, VOF method \citep{MOSHARI2014Num}, and Level-set method \citep{Colicchio2009Expe}.}
	\label{position_convergence_comparison}
\end{figure}
Figure \ref{position_convergence_comparison} 
depicts the time evolution of the vertical position of 
the cylinder obtained with 4 particle resolutions,
to demonstrate the convergence analysis,
and the comparison with results from the literature.
It is observed that, during the initial rising phase, 
the present results are in good agreements with those of experiments and 
previous simulations.  
However, when the cylinder approaches water surface, 
large deviations become apparent,
may be attributed to the different ability to handle 
the "waterfall breaking" and flow separation as discussed in previous sections.
In particular, the results obtained by a previously SPH simulation \citep{Buruchenko2017Vali} and 
the Level-set method \citep{Colicchio2009Expe} 
show a significantly smaller pop-up height compared to the experiment, 
and a significantly smaller increasing slope. 
In contrast, the present results and those obtained by 
the VOF method \citep{MOSHARI2014Num} 
exhibit much closer increasing slope 
and pop-up height compared to the experiment.

\begin{figure}  
	\centering
	\includegraphics[width=\textwidth]{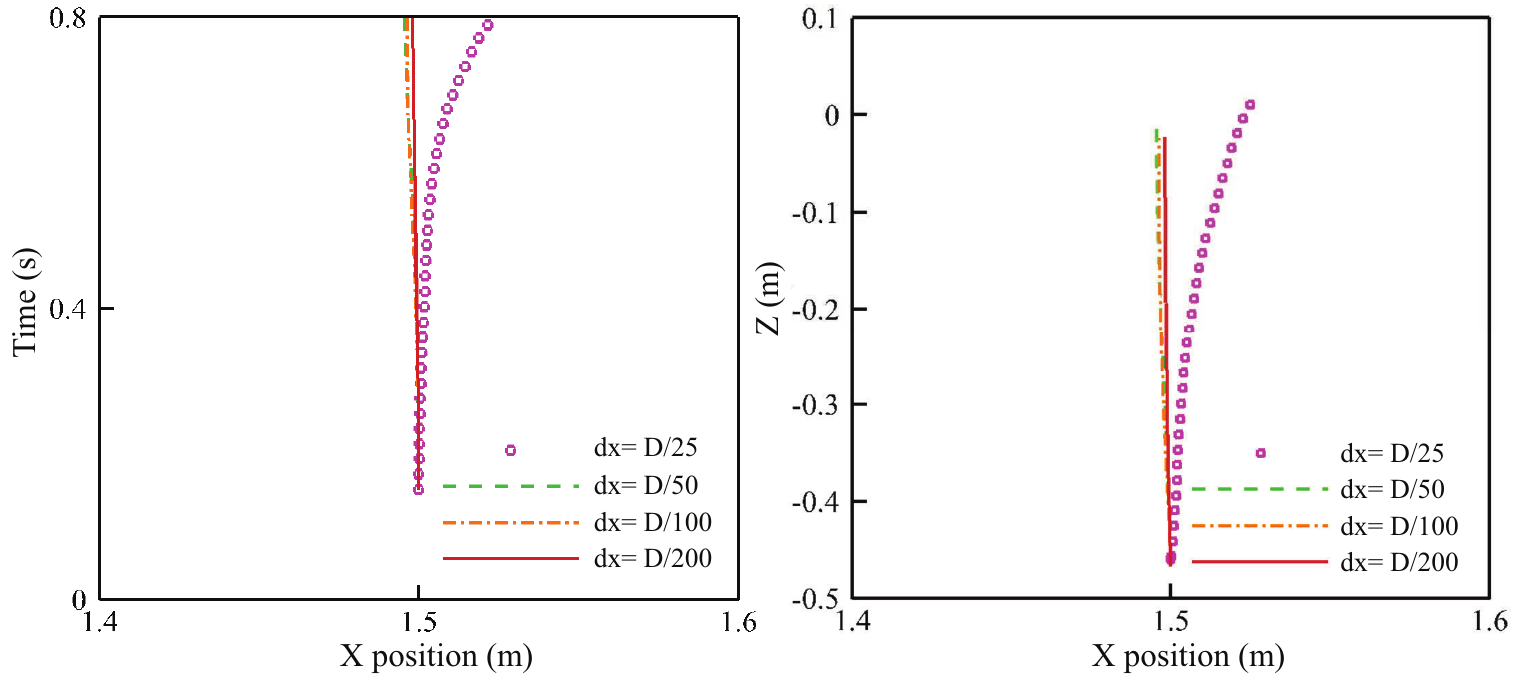}
	\caption{The trajectory of the rising cylinder. Left panel:  the time trace of the lateral position of cylinder center. Right panel: the trajectory of cylinder center in $ X-Z $ plane.}
	\label{lateral_position}
\end{figure}

\begin{figure}
	\centering    
	\includegraphics[width=1\textwidth]{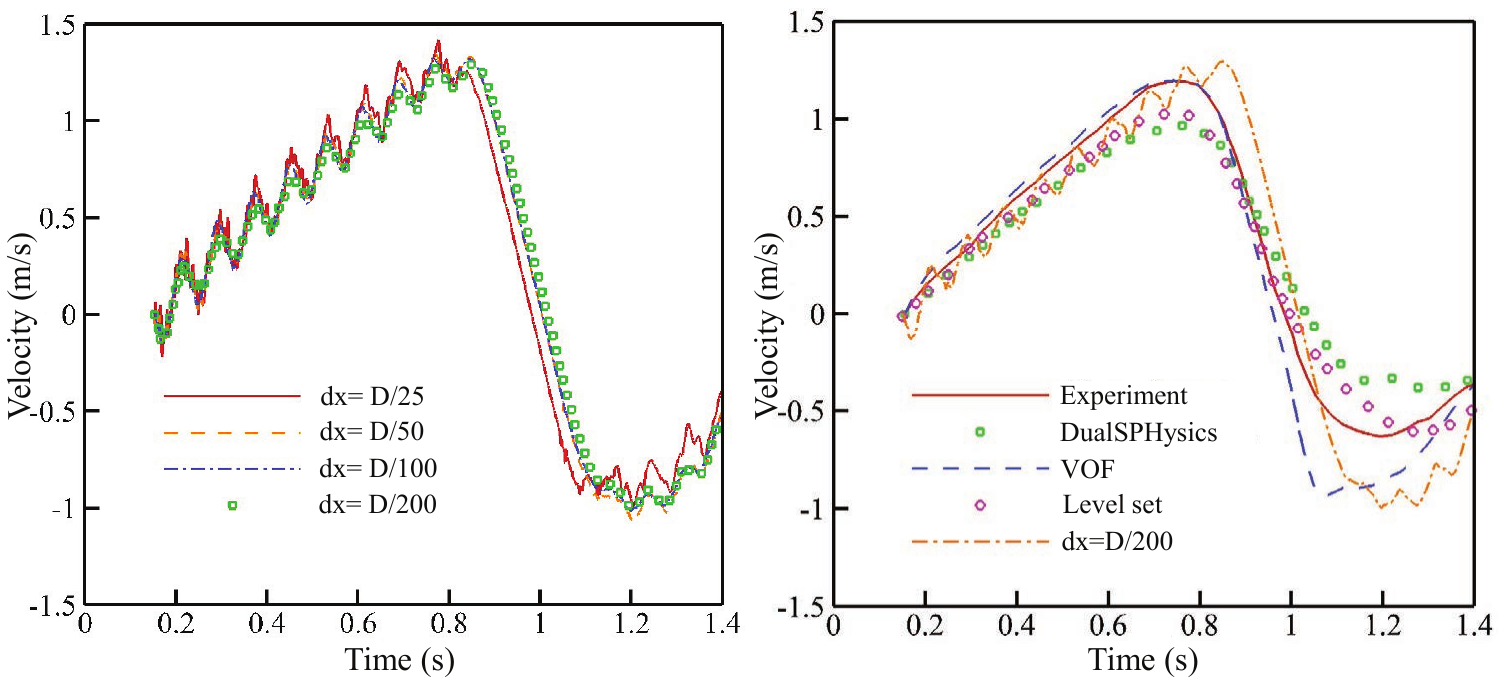}
	\caption{Convergence analysis (left panel) and comparison (right panel) about the vertical velocity of cylinder center. Experiment \citep{Colicchio2009Expe}, DualSPHysics \citep{Buruchenko2017Vali}, VOF method \citep{MOSHARI2014Num}, and Level-set method \citep{Colicchio2009Expe}.}
	\label{velocity_convergence_comparison}
\end{figure}

In previous studies, a sphere with a lower density than water typically vibrates during 
water exit ascent \citep{Newton1999, Schmidt1920, Schmiedel1928, Preukschat1962, 
Kuwabara1983, Veldhuis_2005}, and its ascent is confined to a single vertical plane \citep{Horowitz2008Criti}.
For the 2-D cylinder with a density of $620 kg/m^{3}$ in present simulation, which 
corresponds to the cylindrical cylinder in the experiment, Figure \ref{lateral_position}
illustrates its trajectory during the ascent. The nearly vertical ascent trajectory demonstrates that the rising of the cicurlar cylinder is also confined to a single 
vertical plane.
To further verify the presence of similar vibrations or not, the measured vertical 
position data is temporally derivated. The left panel of Figure
\ref{velocity_convergence_comparison} shows the obtained the time trace of the 
vertical velocity, where an apparent periodic oscillation exists in the vertical 
velocity during the ascent.
In the quantitative comparison of the vertical velocity with the literature, other 
numerical results show some deviations from the experiment during the ascent, but the 
wave crests of the present oscillation curve always fit closely to the filtered 
experimental curve until the moment of "waterfall breaking", as shown in the right panel 
of Figure \ref{velocity_convergence_comparison}.


\section{Conclusion}
\label{Conclusion}
In this study, we propose a diffusive wetting model for water entry/exit based on the WCSPH
method, accounting for the influence of surface wettability on hydrodynamics. The model
includes the diffusive wetting equation, which describes the wetting evolution at the
fluid-solid interface under different surface wettability conditions. Additionally, we
introduce a wetting-coupled spatio-temporal identification approach specifically designed
for interfacial fluid particles. Furthermore, we apply particle regularization to
corresponding interfacial fluid particles to handle various wetting states of the solid.
The proposed model enables accurate simulation of various splashing behaviors in water entry,
due to the consideration of  the effect of surface wettability. It also accurately realizes
the flow separation and spontaneous free-surface breaking in water exit. Moreover, the model
successfully integrates water entry/exit as a complete process in a single numerical simulation. Qualitative and quantitative comparisons with extensive experiments demonstrate the
accuracy, efficiency, and versatility of the proposed model.
As the future work, we plan to further validate the performance of the model by applying it to 
more complex scientific and industrial problems.

\bibliographystyle{jfm}
\bibliography{test}
\end{spacing}
\end{document}